\documentclass[twocolumn,trackchanges]{aastex63}

\usepackage{amsmath}
\usepackage{graphics,color}

\definecolor{darkgreen}{rgb}{0.1, 0.6, 0.2}

\received{XXXX, 2020}
\revised{XXXX, 2020}
\accepted{XXXX}

\newcommand{\be}{\begin{equation}}
\newcommand{\ee}{\end{equation}} 
\newcommand{\lb}{\label}
\newcommand{\OL}{\overline}

\newcommand{\smax}{{\scriptsize\max}}

\newcommand{\bff}{{\bf f}}

\newcommand{\br}{{\bf r}}

\newcommand{\bu}{{\bf u}}

\newcommand{\bx}{{\bf x}}

\newcommand{\bB}{{\bf B}}

\newcommand{\bJ}{{\bf J}}
\newcommand{\bS}{{\bf S}}

\newcommand{\bomega}{\pmb{\omega}}
\newcommand{\bepsilon}{\pmb{\varepsilon}}
\newcommand{\btau}{\pmb{\tau}}

\newcommand{\grad}{{\mbox{\boldmath $\nabla$}}}
\newcommand{\bdot}{{\mbox{\boldmath $\cdot$}}}

\newcommand{\btimes}{{\mbox{\boldmath $\times$}}}

\begin{document}

\title{Scaling of Turbulent Viscosity and Resistivity: Extracting a Scale-dependent Turbulent Magnetic Prandtl Number}

\correspondingauthor{Hussein Aluie}
\email{hussein@rochester.edu}

\author{Xin Bian}
\affiliation{Department of Mechanical Engineering, University of Rochester, NY 14627, USA}
\author{Jessica K. Shang}
\affiliation{Department of Mechanical Engineering, University of Rochester, NY 14627, USA}
 \author{Eric G. Blackman}
\affiliation{Department of Physics and Astronomy, University of Rochester, NY 14627, USA}
 \author{Gilbert W. Collins}
\affiliation{Department of Mechanical Engineering, University of Rochester, NY 14627, USA}
\affiliation{Department of Physics and Astronomy, University of Rochester, NY 14627, USA}
\affiliation{Laboratory for Laser Energetics, University of Rochester, NY 14623, USA}

\author{Hussein Aluie}
\affiliation{Department of Mechanical Engineering, University of Rochester, NY 14627, USA}
\affiliation{Laboratory for Laser Energetics, University of Rochester, NY 14623, USA}

\begin{abstract}

Turbulent viscosity $\nu_t$ and resistivity $\eta_t$ are perhaps the simplest models for turbulent transport of angular momentum and magnetic fields, respectively. The associated turbulent magnetic Prandtl number $Pr_t\equiv \nu_t/\eta_t$ has been well recognized to determine the final magnetic configuration of accretion disks. Here, we present an approach to determining these ``effective transport'' coefficients acting at different length-scales using coarse-graining and recent results on decoupled kinetic and magnetic energy cascades \citep{bian2019decoupled}. 
By analyzing the kinetic and magnetic energy cascades from a suite of high-resolution simulations, we show that our definitions of $\nu_t$, $\eta_t$, and $Pr_t$ have power-law scalings in the ``decoupled range.'' 
We observe that $Pr_t\approx1 \text{~to~}2$ at the smallest inertial-inductive scales, increasing to $\approx 5$ at the largest scales. However, based on physical considerations, our analysis suggests that $Pr_t$ has to become scale-independent and of order unity in the decoupled range at sufficiently high Reynolds numbers (or grid-resolution), and that the power-law scaling exponents of velocity and magnetic spectra become equal. In addition to implications to astrophysical systems, the scale-dependent turbulent transport coefficients offer a guide for large eddy simulation modeling.
\end{abstract}

\keywords{turbulence, Prandtl number, magnetic field, magnetohydrodynamics, large eddy simulation}

\section{Introduction} \label{sec:intro}
Magnetohydrodynamic (MHD) turbulence is central to our understanding of many astrophysical systems, including the solar wind, interstellar medium (ISM), and accretion disks. 

Most of these systems are characterized by very large Reynolds numbers ($Re$). For example, $Re\sim10^5-10^7$ in the cool ISM \citep{elmegreen2004interstellar}, $Re\sim 4\times10^6$ in the solar wind \citep{verma1996nonclassical}, and $Re\sim 10^{14}$ in type Ia supernovae \citep{kuhlen2006carbon}. High-$Re$ turbulent flows involve a wide range of dynamical scales, called the ``inertial-inductive'' range, over which the evolution of the flow and magnetic field are immune from the direct effects of external forcing and microphysical dissipation. Similar to hydrodynamic turbulence, it is widely expected that MHD turbulence over the inertial-inductive range has universal statistics with power-law spectra, although details of such scaling remain a subject of debate \citep{goldreich1995toward,biskamp2003magnetohydrodynamic,zhou2004colloquium,verma2004statistical,verma2019energy,Boldyrev05,schekochihin2020mhd}. While the large scales in a high-$Re$ MHD flow are immune from the \emph{direct} effects of microphysical transport \citep{Aluie17,ZhaoAluie18}, they are indirectly influenced by the microphysics due to the ``catalytic'' role of turbulence via the cascade process, which acts as a bridge between the large and microphysical scales. For example, it is widely believed that turbulence plays an important role in the outward transport of angular momentum in accretion disks for inward mass accretion \citep{balbus1998instability}.

The simplest conceptual framework to think of turbulence is as an effective (or turbulent) viscosity $\nu_t$, which leads to the ``turbulent diffusion'' of angular momentum at scales far larger than viscous scales, and has long shaped our thinking of accretion disk dynamics \citep{ShakuraSunyaev1973}. Similarly, magnetic fields, which are essential for launching and collimating jets \citep{blandford1977electromagnetic, BlandfordPayne1982, Jafari_2018}, can be transported outward by an effective (or turbulent) resistivity $\eta_t$. In this way, the magnetic field configuration in accretion disks may be influenced
by a balance between the inward advection by accretion and the outward diffusion by turbulent resistivity \citep{Lubow+1994,Lovelace+2009,GuanGammie09,FromangStone09,Cao11}. 
This balance between the competing effects of $\nu_t$ and $\eta_t$ is captured by the turbulent magnetic Prandtl number $Pr_t \equiv \nu_t/\eta_t$.  Whether global scale   structures or turbulent stress dominate the overall angular momentum transport is still an open question and  important for determining the budget of thermal vs. non-thermal emission \citep{Blackman+2015}.

For turbulent astrophysical flows, current computing resources are unable to solve all relevant scales. Large eddy simulations (LES) rely on subgrid-scale modeling to represent the small-scale effects on resolved scales \citep{meneveau2000scale,miesch2015large}. \cite{muller2002dynamic,chernyshov2007development,grete2015nonlinear} studied different subgrid-scale (SGS) models. Renormalization group (RG) analysis was used to develop scale-dependent turbulent coefficients \citep{zhou2010renormalization}. However, the studies on MHD scale-dependent turbulent transport coefficients are few compared to hydrodynamic turbulence.

We remind readers that the turbulent magnetic Prandtl number is different from the microscopic magnetic Prandtl number $Pr_m \equiv \nu/\eta$, where $\nu$ is the microscopic viscosity, and $\eta$ is the microscopic resistivity. $Pr_m$ is large in the ISM while being small in stellar interiors and liquid metals \citep{davidson2012ten}. 
Many studies have focused on the effect of $Pr_m$ \citep[e.g.,][]{lesur2007impact,brandenburg2014magnetic,FromangStone09,brandenburg2019reversed}.
The extent to which existing simulations accurately capture the physics of realistic extreme regimes of low and high $Pr_m$ is uncertain.

In this paper we focus on $Pr_t$, not $Pr_m$. Turbulent transport coefficients have been studied both analytically and numerically. Estimates using mixing length theory $\nu_t \approx \eta_t \approx U \ell/3$ (characteristic velocity $U$ and characteristic scale $\ell$) \citep{yousef2003turbulent,kapyla2020turbulent} are consistent to order of magnitude with $\eta_t$ calculated with the test-field method \citep{kapyla2009alpha} and shearing box simulations \citep{snellman2009reynolds}. The quasilinear approximation \citep{kitchatinov1994astron, yousef2003turbulent} and RG analysis \citep{forster1977large,fournier1982infrared,verma2001calculation,verma2001field} suggested that $0.4 < Pr_t < 0.8$. \cite{zhou2002subgrid} developed eddy and backscatter viscosity and resistivity using eddy-damped quasinormal Markovian statistical closure model (EDQNM). 

Numerical studies have traditionally identified ``turbulence'' as fluctuations from a (temporal or ensemble) mean flow, and have typically yielded $Pr_t \approx 1$. 
 \cite{yousef2003turbulent} measured $Pr_t$ from the decaying large-scale fields in forced turbulence simulations. The results showed that $Pr_t$ is near unity and insensitive to $Pr_m$. These simulations were conducted with a fixed small magnetic Reynolds number. Several groups studied the turbulent transport coefficients using shearing box simulations \citep{GuanGammie09, LesurLongaretti09, FromangStone09}. \cite{GuanGammie09} inferred $\eta_t$ from the evolution of an imposed magnetic field perturbation in an already turbulent flow. \cite{LesurLongaretti09} imposed an external magnetic field and defined $\eta_t$ using the electromotive force induced by the field. \cite{FromangStone09} calculated $\eta_t$ from the spatially varying magnetic fields induced by an electromotive term added in the induction equation. $\nu_t$ was defined using Reynolds and Maxwell stress tensors in these studies. Despite different definitions, numerical schemes, and magnetic field configurations among these studies, they all find $Pr_t \approx 1$.
 \cite{kapyla2020turbulent} computed $\nu_t$ using both Reynolds stress and the decay rate of a large-scale field, and $\eta_t$ using the test-field method, where a set of test fields are used to calculate the components of turbulent diffusivity tensors \citep{schrinner2005mean,schrinner2007mean}.
The results suggested that $Pr_t$ increases with increasing Reynolds number and saturates at large Reynolds number with $ 0.8 \le Pr_t\le 0.95$.

Other than the RG and EDQNM analyses, the aforementioned studies did not analyze $\nu_t$ and $\eta_t$ as a function of length-scales, which is not possible from a Reynolds (mean vs. fluctuation) decomposition \citep[e.g.,][]{GuanGammie09,LesurLongaretti09,FromangStone09,kapyla2020turbulent}. Determining the scale dependence of transport coefficients can improve the fidelity with which we
characterize astrophysical turbulence in cohort with its practical application to subgrid scale modeling. For example, if $Pr_t\propto \ell ^{\alpha}$ with $\alpha > 0$, $Pr_t$ grows at larger scales, indicating that the large-scale component of a flow, which is still part of the `fluctuations', feels a stronger $\nu_t$ relative to $\eta_t$.

Our study aims to define and measure $\nu_t$, $\eta_t$, and $Pr_t$ at different scales using the coarse-graining approach \citep{Eyink05,Aluie17} and the eddy-viscosity hypothesis \citep{boussinesq1877essai}. Our analytical and numerical results show power-law scaling of the turbulent transport coefficients in the ``decoupled range'' over which the kinetic and magnetic cascades statistically decouple and become conservative \citep{bian2019decoupled}. 

\section{Methodology}\label{sec:method}

\subsection{Coarse-grained energy equations}

We analyze the incompressible MHD equations with a constant density $\rho$
\begin{gather}
\partial _t \textbf u+ ( \textbf u\bdot \grad) \textbf u   =  - \grad p+ \bJ \btimes \bB +\nu \nabla^2 \bu+ \bff,\label{velmhd}\\
\partial _t \textbf B =  \grad\btimes(\textbf u\btimes \textbf B)+\eta\nabla^2 \textbf B, \label{eq:magmhd}\\
\grad\bdot\bu =  \grad\bdot\bB=0,  \label{eq:incompressible}
\end{gather}
where $\bu$ is the velocity, and $\bB$ is the magnetic field normalized by $\sqrt{4\pi\rho}$ to have Alfv\'en (velocity) units. $p$ is pressure, $\bJ=\grad\btimes\bB$ is (normalized) current density, $\textbf f$ is external forcing, $\nu$ and $\eta$ are microscopic viscosity and resistivity, respectively. 

We use the coarse-graining method to analyze the flow and define the turbulent magnetic Prandtl number. A coarse-grained field in $n$-dimensions $\OL f_\ell(\bx) = \int d^n\br\, G_\ell(\bx - \br) f(\br)$ contains modes at length-scales greater than $\ell$, where $G_\ell(\br)\equiv \ell^{-n} G(\br/\ell)$ is a normalized kernel with its main weight in a ball of diameter $\ell$. The coarse-grained MHD equations for $\overline{\textbf u}_{\ell}$, $\overline{\textbf B}_{\ell}$, and the quadratic MHD invariants were shown by \cite{Aluie17}. Hereafter, we drop subscript $\ell$ when possible.

The coarse-grained kinetic energy (KE) and magnetic energy (ME) density balance (at scales $>\ell$) are,
\begin{align}
\partial_t&(\frac{|\overline{\textbf {u}}|^2}{2})+\grad\bdot[\cdots]\nonumber\\
&=-\OL\Pi^u_{\ell}-\overline S_{ij} \overline B_i \overline B_j-2\nu| \overline {\bS}|^2+\OL\bff\bdot\OL\bu,\label{kineticenergy} \\
\partial_t&(\frac{|\OL{\bB}|^2}{2})+\grad\bdot[\cdots]\nonumber\\
&=-\OL\Pi^b_{\ell}+ \overline S_{ij} \overline B_i \overline B_j-\eta | \overline {\bJ}|^2, \label{mageticenergy}
\end{align}
where $\grad\bdot[\cdots]$ denotes spatial transport terms, $\bS=(\grad\bu + \grad\bu^{T})/2$ is the strain-rate tensor, $\OL\bff\bdot\OL\bu$ is the energy injection rate at forcing scale $\ell_f = 2\pi/k_f$ ($k_f$ are the modes of the forcing $\bff$). 
 Microscopic dissipation terms $\nu| \overline {\bS}|^2$ and 
$\eta | \overline {\bJ}|^2$ are mathematically guaranteed \citep{Aluie17,Eyink18} and numerically demonstrated \citep{ZhaoAluie18, bian2019decoupled} to be negligible at scales $\ell\gg{\left(\ell_\nu,\ell_\eta\right)}$, where $\ell_\nu$ and $\ell_\eta$ are the viscous and resistive length scales, respectively.

The KE cascade term $\OL\Pi^u_{\ell}\equiv-\OL\bS_\ell\mathbin{:}\OL\btau_\ell$ in eq. \eqref{kineticenergy} quantifies the KE transfer \emph{across} scale $\ell$, where $\OL\tau_{ij} \equiv \tau_{\ell}(u_i, u_j)-\tau_{\ell}(B_i, B_j)$ is the sum of subscale Reynolds and Maxwell stresses generated by scales $<\ell$ acting on the large-scale strain $\OL{S}_{ij}$, resulting in ``turbulent viscous dissipation'' to scales $<\ell$. 
Subscale stress is defined as $\tau_{\ell}(f, g)=\OL{\left(f g\right)}_{\ell}-\overline f_{\ell} \overline g_{\ell}$ for any two fields $f$ and $g$. Similarly, ME casade term $\OL\Pi^b_{\ell}\equiv-\OL{\bJ}_\ell\bdot \OL\bepsilon_\ell$ in eq. \eqref{mageticenergy} quantifies the ME transfer \emph{across} scale $\ell$, where the subscale electromotive force (EMF) $\OL\bepsilon_\ell \equiv \OL{\bu\btimes\bB} - \OL{\bu}\btimes\OL{\bB}$ is (minus) the electric field generated by scales $<\ell$ acting on the large-scale current $\OL\bJ=\grad\btimes\OL\bB$, resulting in ``turbulent Ohmic dissipation'' to scales $<\ell$. Both $\OL\Pi^u_{\ell}$ and $\OL\Pi^b_{\ell}$ appear as sinks in the energy budgets of large scales $>\ell$ and as sources in the energy budgets of small scales $<\ell$ \citep{Aluie17}. 

Term $\overline S_{ij} \overline B_i \overline B_j$ quantifies KE-to-ME conversion at all scales $>\ell$ and appears as a sink in eq. \eqref{kineticenergy} and a source in eq. \eqref{mageticenergy}. \cite{bian2019decoupled} showed that $\langle\overline S_{ij} \overline B_i \overline B_j\rangle$ ($\langle ... \rangle$ denotes a spatial average) is a large-scale process, which only operates at the largest scales in the inertial-inductive range (which was called the ``conversion range'') and vanishes at intermediate and small scales in the inertial-inductive range (which was called the ``decoupled range''). In the decoupled range, $\langle\OL\Pi^u_{\ell}\rangle$ and $\langle\OL\Pi^b_{\ell}\rangle$ become constant as a function of scale (i.e., scale-independent). The observation of constant KE and ME fluxes $\langle\OL\Pi^u_{\ell}\rangle$ and $\langle\OL\Pi^b_{\ell}\rangle$ is important since it indicates separate conservative cascades of each of KE and ME, which arises asymptotically at high Reynolds number regardless of forcing, external magnetic field, and microscopic magnetic Prandtl number.

\subsection{Scaling of turbulent transport coefficients}
Oftentimes, turbulence is modeled as a diffusive process via effective (or turbulent) transport coefficients. For example, mixing length or eddy viscosity models represent the subscale stress $\OL\btau_\ell$, due to scales $<\ell$, as $\OL{\tau}_{ij} = - 2 \nu_t^\bx  \overline S_{ij}$, where $\nu_t^\bx$ is a turbulent viscosity\footnote{Strictly speaking, the eddy viscosity definition is $\mathring{\OL{\tau}}_{ij} = - 2 \nu_t \overline S_{ij}$, where $\mathring{\OL{\tau}}_{ij}=\OL{\tau}_{ij} - \delta_{ij}\OL{\tau}_{kk}/3$ is the traceless part of the stress. For our incompressible flow analysis here, which is based on the energy flux, this distinction does not matter.} (e.g., \cite{pope2001turbulent} and references therein). {Similarly, the subscale EMF can be modeled as $\OL{\bepsilon} = - \eta_t^\bx \OL\bJ + \alpha\OL{\bB}$, where the $\eta_t^\bx \OL\bJ$ term models the subscales as turbulent resistive diffusion \citep{miesch2015large} and the $\alpha\OL{\bB}$ term is the $\alpha$-effect of dynamo theory \citep{Moffatt+1978}. The $\alpha$-effect is expected to play a role in flows where the driving mechanism is helical. To simplify our analysis and the presentation of our approach, we shall neglect the $\alpha\OL{\bB}$ term and assume that the subscale EMF can be modeled solely as Ohmic diffusion, $\OL{\bepsilon} = - \eta_t^\bx \OL\bJ$.} Note that $\nu_t^\bx(\bx, t, \ell)$ and $\eta_t^\bx(\bx, t, \ell)$ are generally functions of space $\bx$, length scale $\ell$, and time.

A main goal of this paper is extracting the turbulence transport coefficients, $\nu_t^\bx$ and $\eta_t^\bx$, as a function of length-scale. However, we do not pursue a phenomenological analysis similar to that of \cite{smagorinsky1963general} or of a mixing length framework \citep{tennekes1972first} in part because we lack a consensus MHD turbulence theory analogous to that of \cite{kolmogorov1941local}. To achieve our goal, we shall instead analyze the \emph{energy budgets} resultant from the eddy viscosity model. Within our coarse-graining framework, this is equivalent to having the rate of energy cascading to scales smaller than $\ell$ equal a turbulent dissipation acting on scales $>\ell$:
\begin{eqnarray}
2\nu_t \langle |\OL \bS_\ell|^2 \rangle &\equiv& \langle \OL\Pi^u_{\ell} \rangle ,\label{eq:pi_u_nu}  \\
\eta_t \langle |\OL{\bJ}_\ell|^2 \rangle &\equiv& \langle \OL\Pi^b_{\ell} \rangle  \label{eq:pi_b_eta}.
\end{eqnarray}
These two relations are definitions for $\nu_t$ and $\eta_t$. Note that unlike in relation $\OL{\tau}_{ij} = - 2 \nu_t^\bx  \overline S_{ij}$, the turbulence transport coefficients in eqs. \eqref{eq:pi_u_nu}-\eqref{eq:pi_b_eta} are defined using \emph{scalar} quantities $\langle \OL\Pi^u_{\ell}\rangle$, $\langle \OL\Pi^b_{\ell} \rangle$, $\langle |\OL \bS_\ell|^2 \rangle$, and $\langle |\OL{\bJ}_\ell|^2 \rangle$. For homogeneous turbulence considered in this study, we rely on spatial averages, $\langle\dots\rangle$, rendering $\nu_t$ and $\eta_t$ independent of location $\bx$ but still a function of scale $\ell$. 

Consistent with the eddy viscosity hypothesis, eq. \eqref{eq:pi_u_nu} (eq. \eqref{eq:pi_b_eta}) models the kinetic (magnetic) energy cascading from scales $>\ell$ to smaller scales as effectively being dissipated by a turbulent viscosity (resistivity). 
From these, we can also extract a scale-dependent turbulent magnetic Prandtl number,  
\begin{gather}
Pr_t\equiv\nu_t/\eta_t = \frac{\langle\OL \Pi^u_{\ell}\rangle }{\langle\OL \Pi^b_{\ell}\rangle }\frac{\langle |\OL{\bJ}_\ell|^2 \rangle}{\langle2 |\OL \bS_\ell|^2 \rangle}. \label{eq:pr_def}
\end{gather}

What power-law scaling can we expect these turbulent transport coefficients to have?
It is possible to relate $\nu_t$ and $\eta_t$ to energy spectra. Indeed, the space-averaged turbulent dissipation can be expressed in terms of energy spectra:
\begin{gather}
\langle \OL\Pi^u_{\ell} \rangle =  2\nu_t  \langle  |\OL {\bS}_{\ell}|^2 \rangle = 2 \nu_t \int_{0}^{k} k'^2 E^{u}(k') dk', \label{eq:nu_spec_u} \\
\langle \OL\Pi^b_{\ell} \rangle  = \eta_t \langle| \OL {\bJ}_{\ell}|^2 \rangle = 2 \eta_t \int_{0}^{k} k'^2 E^{b}(k') dk', \label{eq:eta_spec_b}
\end{gather}
where $E^{u}(k)$ ($E^{b}(k)$) is the kinetic (magnetic) energy spectrum with (dimensionless) wavenumber $k=L/\ell$ for a periodic domain of size $L$.

The scaling of spectra in turn are related to the scaling of velocity and magnetic field increments \citep{Aluie17}
\begin{gather}
\delta u(\ell) \propto \ell^{\sigma_u},
\delta B(\ell) \propto \ell^{\sigma_b}, \label{eq:increment_scale}
\end{gather}
where increment $\delta f(x;  \ell) = f(x +  \ell) - f(x)$ (see details in \cite{Eyink05,AluieEyink10,Aluie17}). From eq. \eqref{eq:increment_scale}, the kinetic and magnetic energy spectra scale as
\begin{gather}
E^{u}(k)\propto k ^{-2\sigma_u - 1},
E^{b}(k)\propto k ^{-2\sigma_b - 1}. \label{eq:spec_scale}
\end{gather}
The relation between increments and spectra does not make any assumptions about the specific exponent values, only that they are $ \sigma_{u, b} < 1$ \citep{SadekAluie18}. Scaling exponents $\sigma_{u}$ and $\sigma_{b}$ are a measure of smoothness of the velocity and magnetic fields, respectively (see Fig. 1 and related discussion in \cite{Aluie17}). A value of $\sigma=1$ indicates that the field is very smooth (e.g., of a laminar flow) with a spectrum decaying as $k^{-3}$ or steeper. Canonical hydrodynamic turbulence has intermediate smoothness, with $\sigma_u=1/3$ according to \cite{kolmogorov1941local} (K41). The larger is the value of $\sigma$, the smoother is the field. 

For sufficiently high Reynolds number flows, \cite{bian2019decoupled} showed that each of $\langle\OL \Pi^u_{\ell}\rangle$ and $\langle\OL \Pi^b_{\ell}\rangle $ become constant, independent of scale in the decoupled range. From definitions \eqref{eq:pi_u_nu}-\eqref{eq:pr_def}, and considering the scaling relations discussed above, we can infer that the turbulent transport coefficients vary with scale as follows:
\begin{align}
\nu_t \propto k^{-2(1-\sigma_u)}, 
\eta_t  \propto k^{-2(1-\sigma_b)},
Pr_t \propto k^{-2(\sigma_b-\sigma_u)},\label{eq:scaling_decopled}
\end{align}
for scales $k$ in the decoupled range. It is possible to obtain scaling relations \eqref{eq:scaling_decopled} from either the scaling of spectra in eqs. \eqref{eq:nu_spec_u}-\eqref{eq:eta_spec_b}, or the scaling of coarse-grained strain and current,
$|\overline{\bS}_{\ell}| \sim \delta u(\ell) / \ell$ and $|\overline{\bJ}_{\ell}| \sim \delta B(\ell) / \ell$ \citep{eyink2013flux,Aluie17}. Eq. \eqref{eq:scaling_decopled} highlights that $Pr_t$ is independent of scale only if $\sigma_u=\sigma_b$.  

Regardless of the specific value, and consistent with existing MHD turbulence phenomenologies, we expect that $\sigma_{u,b}<1$. Indeed, a $\sigma_{u,b}\ge1$ would correspond to a smooth flow that is inconsistent with the qualitative expectation of a `rough' or `fractal' turbulent flow. Therefore, relations \eqref{eq:scaling_decopled} indicate that $\nu_t$ and $\eta_t$ decay as $\ell\to0$ (or $k\to\infty$). This is consistent with physical expectations since the `eddies' effecting the turbulent transport become weaker at smaller scales, yielding smaller transport coefficients. 

We highlight a technical, albeit important  aspect of scaling relations \eqref{eq:scaling_decopled}.
Our coefficients seem to scale with the inverse of coarse-grained strain and current magnitudes, $\nu_t\sim |\overline{\bS}_{\ell}|^{-2}\sim \ell^{2-2\sigma_u}$ and $\eta_t\sim |\overline{\bJ}_{\ell}|^{-2}\sim \ell^{2-2\sigma_b}$, but do not appear to depend on the subscale stress and EMF, $\OL\btau_\ell$ and $\OL\bepsilon_\ell$, respectively. At face value, this result seems counter-intuitive wherein $\sigma\to1$ associated with smoother fields and weaker `eddies' leads to an increase rather than a drop in the turbulent coefficient values in eq. \eqref{eq:scaling_decopled}. However, a key assumption in arriving at relations \eqref{eq:scaling_decopled} is that fluxes $\langle\OL \Pi^u_{\ell}\rangle$ and $\langle\OL \Pi^b_{\ell}\rangle $ are constant, independent of scale. For scale-independent fluxes to be established, consistent with a persistent cascade to arbitrarily small scales (in the $Re\to\infty$ limit), $\sigma_{u}$ and $\sigma_{b}$ have to take on fixed values that are yet to be determined and agreed upon by the community. If $\sigma_{u,b}$ were to be somehow increased above those values, the cascade would shut down (fluxes would decay with $k$) before carrying the energy all the way to dissipation scales \citep{Aluie17}. For scale-dependent fluxes, relations \eqref{eq:scaling_decopled} have to be modified to also include the scaling of $\langle\OL \Pi^u_{\ell}\rangle$ and $\langle\OL \Pi^b_{\ell}\rangle $ (see \cite{Aluie17} for details).

{To infer the scaling of turbulence transport coefficients, the approach we adopt in this paper circumvents using values of $\sigma_{u}$ and $\sigma_{b}$ (in the asymptotic $Re\to\infty$ limit) from a specific MHD phenomenology--whether it exists or not-- by relying on results from \cite{bian2019decoupled} of scale-independent fluxes $\langle\OL \Pi^u_{\ell}\rangle$ and $\langle\OL \Pi^b_{\ell}\rangle$.}

Under K41 scaling $\sigma_u = 1/3$ \citep{kolmogorov1941local}, our scaling of $\nu_t \propto \ell^{2-2\sigma_u}  \propto \ell^{4/3}$ is equivalent to that from mixing length theory $\nu_t = \ell^2 |\overline{\bS}| \propto \ell^{\sigma_u+1}\propto \ell^{4/3}$ \citep{smagorinsky1963general}. 
Our analysis is also compatible with different scaling theories and observations in MHD turbulence \citep{iroshnikov1963turbulence,Kraichnan65,goldreich1995toward,Boldyrev05,Boldyrev06,boldyrev2009spectrum,Boldyrevetal11}. For example, solar wind observations \citep{podesta2007spectral,borovsky2012velocity} suggest that $E^u(k)\sim k^{-3/2}$ for the kinetic energy spectrum, corresponding to $\delta u(\ell) \sim \ell^{1/4}$, and $E^b(k)\sim k^{-5/3}$ for the magnetic energy spectrum, corresponding to $\delta B(\ell) \sim \ell^{1/3}$, yield
\begin{align}
\nu_t \sim k^{-3/2}, \eta_t \sim k^{-4/3}, Pr_t \sim k^{-1/6}, \label{eq:scaling_solar}
\end{align}
for $k$ in the decoupled scale-range.

\subsection{Alternate measure of the coefficients}
Instead of analyzing the energy budgets to determine $\nu_t$, $\eta_t$, $Pr_t$ and their scaling, we can alternatively focus on the budgets for vorticity and current. Similar to eqs. \eqref{kineticenergy}-\eqref{mageticenergy}, we can derive the budgets 
\begin{eqnarray}
\partial_t(\frac{|\OL{\bomega}|^2}{2}) + \grad\bdot[\dots] = \dots - \OL{Z}_\ell \lb{eq:enstrophy}\\
\partial_t(\frac{|\OL{\bJ}|^2}{2}) + \grad\bdot[\dots] = \dots -  \OL{Y}_\ell \lb{eq:Current}
\end{eqnarray}
Here, $\OL{Z}_\ell =   \OL\bomega\bdot \grad\btimes(\grad\bdot\OL\btau) $ and $\OL{Y}_\ell =-\OL\bJ\bdot \grad\btimes\grad\btimes\OL\bepsilon$ are the only ``scale-transfer'' terms in the coarse-grained eqs. \eqref{eq:enstrophy}-\eqref{eq:Current} involving the interaction of subscale terms $\OL\btau_\ell$ and $\OL\bepsilon_\ell$ with large-scale quantities (here, $\OL\bomega$ and $\OL\bJ$) to cause transfer across scale $\ell$. From the models $\OL\btau_\ell = -2\nu_t\OL\bS_\ell$ and $\OL\bepsilon_\ell = -\eta_t\OL\bJ_\ell$, we have alternate definitions for the turbulent transport coefficients:
\begin{eqnarray}
\nu_t \equiv \frac{1}{4} \frac{\langle\OL{Z}_\ell\rangle}{\langle|\grad\bdot\OL\bS_\ell|^2\rangle} \lb{eq:TurbViscDef2}\\
\eta_t \equiv \frac{\langle\OL{Y}_\ell\rangle}{\langle|\grad\btimes\OL\bJ_\ell|^2\rangle} \lb{eq:TurbResistDef2}
\end{eqnarray}
Note that unlike energy, vorticity and current
density are not ideal invariants and, therefore, do not undergo a cascade in the manner energy does. Yet, to the extent $\nu_t$ and $\eta_t$ are able to capture the subscale physics embedded in $\OL\btau_\ell$ and $\OL\bepsilon_\ell$, it is reasonable to expect that the turbulent transport coefficients are consistent with the budget of any quantity derived from the underlying dynamics. 

In Fig.~\ref{fig:vis_res_pr}, we compare $\nu_t$ and $\eta_t$ when calculated from eqs. \eqref{eq:TurbViscDef2}-\eqref{eq:TurbResistDef2} to those obtained from the energy budgets in eqs. \eqref{eq:pi_u_nu}-\eqref{eq:pi_b_eta}. We find that the two definitions yield fairly similar results with slight quantitative differences. This consistency lends support to our approach of using the energy budgets to calculate $\nu_t$ and $\eta_t$ (eqs. \eqref{eq:pi_u_nu}-\eqref{eq:pi_b_eta}) and make inferences about turbulent diffusion or dissipation of quantities other than energy.

\subsection{Implications to subgrid modeling}
It is almost always the case that astrophysical systems of interest are at sufficiently high Reynolds numbers (both magnetic and hydrodynamic) that it is impossible to simulate the entire dynamic range of scales that exist \citep{miesch2015large}. In practice, most simulations are either explicit or implicit Large eddy simulations (LES), resolving only the large-scale dynamics \citep{meneveau2000scale}. The former include explicit terms in the equations being solved that model the unresolved subgrid physics, whereas the latter rely on the numerical scheme to act as a de facto model for such missing physics. Our analysis here can offer guidance for tuning the turbulent coefficients when conducting explicit Large Eddy Simulations using eddy diffusivity models. It can also offer us insight into whether relying on a similar scheme and grid for simulating both the momentum and magnetic fields is justified. 

In the inertial-inductive range, using eq. \eqref{eq:pi_u_nu}, $|\overline{\bS}_{\ell}| \sim \delta u(\ell) / \ell$, and the Ansatz \citep{Aluie17} 
\be
\delta u(\ell) \propto u_{rms} \left(\frac{\ell}{L}\right) ^{\sigma_u},
\lb{eq:AnsatzU}\ee 
where $u_{rms} = \langle |{\bu}|^2\rangle^{1/2}$,  
$L$ is a characteristic large scale such as the integral scale or that of the domain, and we ignore intermittency corrections, we then have
\begin{align}
\nu_{t} &=\frac{\langle \OL \Pi^u_{\ell} \rangle}{\langle2|\overline{\bS}_{\ell}|^{2}\rangle}
\sim\frac{\langle\OL \Pi^{u}_{\ell}\rangle \ell^{2}}{|\delta u|^{2}}
\sim \frac{\langle \OL \Pi^u_{\ell} \rangle} {u_{rms}^2/L^2} \quad\left(\frac{\ell}{L}\right)^{2-2\sigma_{u}}.
\end{align}

In an LES with grid spacing $\Delta x$, the turbulent viscosity accounting for subgrid scales should be evaluated at a coarse-graining scale $\ell_c=L/k_c$
proportional to $\Delta x$ \citep{pope2001turbulent}, where $k_c=L/\ell_c$ is a {dimensionless} cutoff wavenumber:
\begin{align}
\nu_{t}(k=k_c) = \frac{\langle \OL \Pi^u_{\ell} \rangle} {2 C_u^2 u_{rms}^2/L^2} \left(\frac{1}{k_c}\right)^{2-2\sigma_{u}},\label{eq:les_nu}
\end{align}
for $\left(\ell_\nu,\ell_\eta\right)\ll\ell_c\ll L$, where dimensionless constant $C_u$ defined as the proportionality factor of the relation
\begin{align}
\langle|\overline{\bS}_{\ell}|^{2}\rangle^{1/2} = C_u \frac{u_{rms}}{L} \left(\frac{\ell}{L}\right)^{\sigma_{u}-1}.
\end{align}
Figure \ref{fig:cucb} in the Appendix shows that $C_u$ is indeed a proportionality constant that is scale-independent within the decoupled range, taking on values from 2 to 5 in various simulated flows.

Similarly, the turbulent resistivity at the cutoff wavenumber is
\begin{align}
\eta_{t}(k=k_c) = \frac{\langle \OL \Pi^b_{\ell} \rangle} {C_b^2 B_{rms}^2/L^2} \left(\frac{1}{k_c}\right)^{2-2\sigma_{b}},\label{eq:les_mu}
\end{align}
where {$B_{rms} = \langle|\bB-\bB_0|^2\rangle^{1/2}$} ($\bB_0$ is the uniform external magnetic field), and dimensionless constant $C_{b}$ defined as 
\begin{align}
\langle|\overline{\bJ}_{\ell}|^{2}\rangle^{1/2} = C_b \frac{B_{rms}}{L} \left(\frac{\ell}{L}\right)^{\sigma_{b}-1}.
\end{align}
{Fig. \ref{fig:cucb} in the Appendix shows that $C_b$ is indeed a proportionality constant that is scale-independent within the decoupled range, taking on values from 10 to 15 in various simulated flows.}

If the grid is sufficiently fine to resolve some of the scales in the decoupled range, then eqs. \eqref{eq:les_nu},\eqref{eq:les_mu} simplify to
\begin{align}
\nu_{t}(k=k_c) &=  \frac{\varepsilon_u} {2C_u^2 u_{rms}^2/L^2}  \left(\frac{1}{k_c}\right)^{2-2\sigma_{u}},\label{eq:les_nu_decoupled}\\
\eta_{t}(k=k_c) &= \frac{\varepsilon_b} {C_b^2 B_{rms}^2/L^2} \left(\frac{1}{k_c}\right)^{2-2\sigma_{b}}\label{eq:les_mu_decoupled},
\end{align}
with scale-independent fluxes $\langle\OL\Pi^{u}_\ell\rangle=\varepsilon_u$ and $\langle\OL\Pi^{b}_\ell\rangle=\varepsilon_b$. These are the KE and ME cascade rates, which were found by \cite{bian2019decoupled} to reach equipartition in the decoupled range,  $\varepsilon_u = \varepsilon_b = \varepsilon/2$, half the total energy cascade rate, $\varepsilon$.

Eqs. \eqref{eq:les_nu},\eqref{eq:les_mu} (and eqs. \eqref{eq:les_nu_decoupled},\eqref{eq:les_mu_decoupled}) connect the scaling of turbulent transport coefficients with the scaling of velocity and magnetic spectra, and are compatible with different MHD scaling theories. For example, if $E^u(k)\sim k^{-5/3}$ (corresponding to $\delta u(\ell) \sim \ell^{1/3}$) as in the theory by \cite{GoldreichSridhar95}, eq. \eqref{eq:les_nu} reduces to the turbulent viscosity model of \cite{verma2001calculation} derived from an RG analysis (see also \cite{Verma2004LES}).

\begin{table}[ht]
\caption{Simulation parameters. ABC (helical) and Taylor-Green (non-helical) forcing are applied at wavenumber $k_f$. $B^\smax_k=\sqrt{\max_k[E^b(k)]}$ is at the magnetic spectrum's [$E^b(k)$] peak. Each simulation set includes runs with the same parameters except grid resolution (Reynolds numbers). $N^3_{max}$ denotes the highest resolution in each set. Subscripts a, b, c, and d denote resolution of $256^3$, $512^3$, $1024^3$, and $2048^3$, respectively. More details are given in Table \ref{Tbl:DetailedParameters} in Appendix.}
\begin{center}
\begin{tabular}{lccccccc}
\hline
\hline
Run            & Forcing & $k_f$ & $Pr_m$  & $|\bB_0|/B^\smax_k$ & $N_{max}^3$\\ \hline
   $\rm I$     & ABC        & 2    & 1     & 0   &$1024^3$             \\
   $\rm II$    & ABC        & 2    & 1     & 10  &$1024^3$               \\
   $\rm III$   & TG         & 1    &  1     & 0   &$1024^3$     \\
   $\rm IV$    & TG       & 1    & 2     & 0   &$1024^3$     \\ 
   $\rm IV$($Pr_m$=0.1)   & TG       & 1    &  0.1     & 0   &$512^3$     \\
   $\rm IV$($Pr_m$=5)   & TG         & 1    &  5     & 0   &$512^3$     \\
   $\rm IV$($Pr_m$=10)   & TG         & 1    &  10     & 0   &$512^3$     \\
   $\rm V$     & ABC        & 2    & 1     & 2   &$2048^3$     \\ 
\hline
\end{tabular}
\end{center}
\label{Tbl:Simulations}
\end{table}

\section{Numerical Results}\label{sec:result}

We conduct pseudo-spectral direct numerical simulations (DNS) of MHD turbulence using hyperdiffusion with grid resolutions up to $2048^3$. Simulation parameters are summarized in Table \ref{Tbl:Simulations} (see details in Table \ref{Tbl:DetailedParameters} in Appendix). To discern trends in the high-$Re$ asymptotic limit, each set of simulations is run under the same parameters but at different grid resolutions (Reynolds numbers). Our flows are driven with either non-helical forcing (Runs $\rm III$ and $\rm IV$) or helical forcing (Runs $\rm I$,  $\rm II$, and $\rm IV$). {Since we do not account for  the $\alpha$-effect when modeling the turbulent EMF, $\OL{\bepsilon}$, which may be important in helically-driven flows, we focus on results from Runs $\rm III$ and $\rm IV$ in the main text while those driven with helical forcing (Runs $\rm I$, $\rm II$, and $\rm IV$) are shown in Appendix for completeness. We note that all simulations yield remarkably similar results, regardless of the type of forcing.}

\begin{figure*}[!htb]
\gridline{\fig{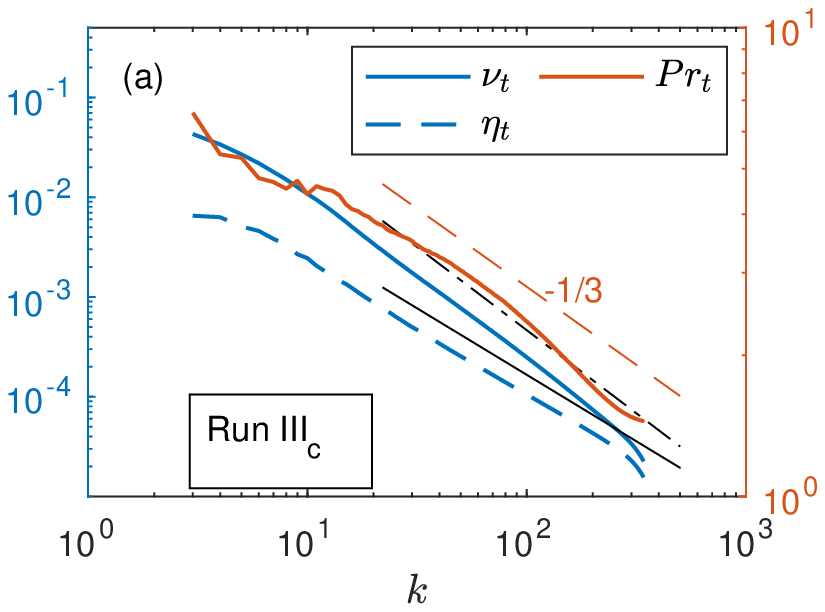}{0.35\textwidth}{}
          \fig{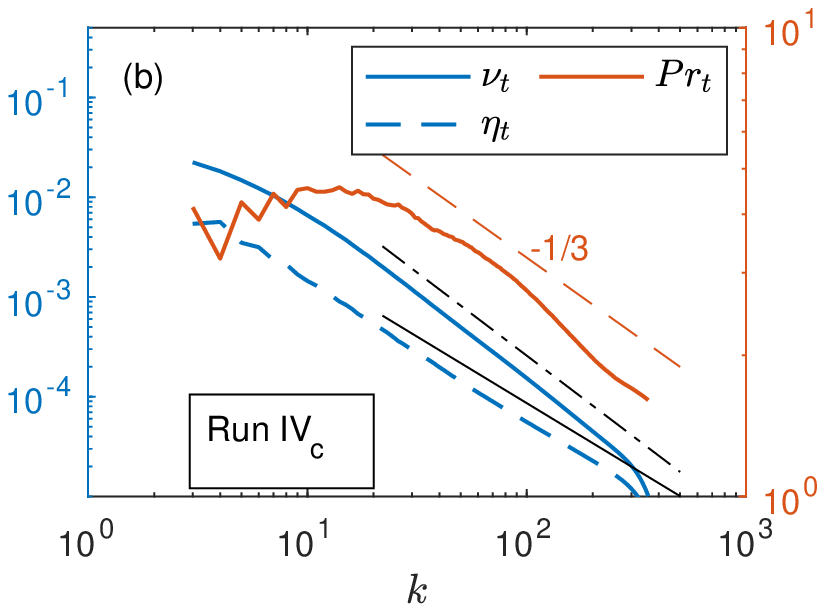}{0.35\textwidth}{}
          }
\gridline{\fig{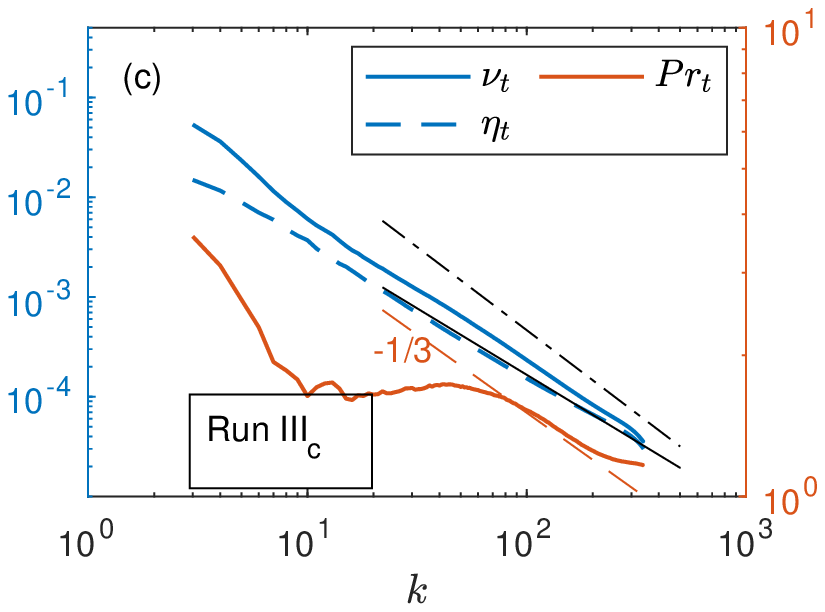}{0.35\textwidth}{}
          \fig{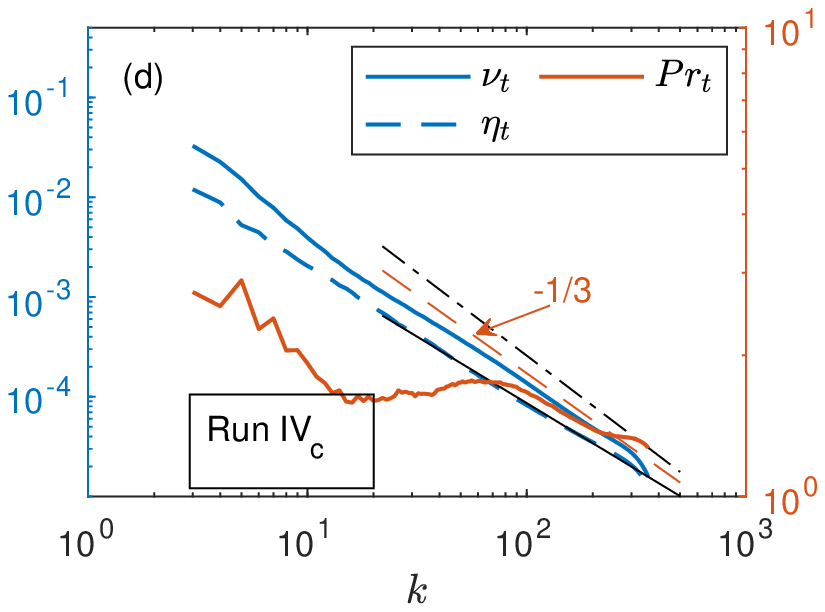}{0.35\textwidth}{}
          }
\caption{Panels (a)-(b) show $\nu_t$, $\eta_t$, and $Pr_t$ calculated using their respective definitions in eqs. \eqref{eq:pi_u_nu}-\eqref{eq:pr_def}, at different scales $k = L/\ell$. Panels (c)-(d) show an alternate calculation of $\nu_t$ and $\eta_t$ from eqs. \eqref{eq:TurbViscDef2}-\eqref{eq:TurbResistDef2}. We use the highest resolution runs of Run III and IV (Taylor-Green forcing) in Table \ref{Tbl:Simulations}. Three reference lines with a slope of -1/3, -5/3 (black dash-dotted), and -4/3 (black solid) are added. Note the reference line of -1/3 and $Pr_t$ use the RIGHT $y$-axis, while others use the LEFT $y$-axis. Scales $<\ell_d$ are not shown. Simulations with helical forcing are shown in Fig. \ref{fig:vis_res_pr-abc} in Appendix. 
}
\label{fig:vis_res_pr}
\end{figure*}

In our simulations, we observe a scaling of $E^{u}(k)\sim k^{-4/3}$ in Run $\rm I_c$, $\rm III_c$, and $\rm IV_c$, and $E^{u}(k)\sim k^{-3/2}$ in Run $\rm II_c$ and $\rm V_d$ (see Figs. \ref{fig:spec_u},\ref{fig:spec_b} in Appendix), corresponding to scaling exponents of $\sigma_u=1/6$ and $\sigma_u=1/4$, respectively. $E^{b}(k)\sim k^{-5/3}$ in all runs (at the highest resolution), corresponding to scaling exponents of $\sigma_b=1/3$. The $E^{b}(k)$ scaling is consistent with that reported in solar wind studies \citep{podesta2007spectral,borovsky2012velocity}. 
The $E^{u}(k)\sim k^{-4/3}$ scaling is consistent with that reported by \cite{grete2020matter} using the code K-Athena \citep{stone2020athena++,grete2020k}, and is slightly shallower than $k^{-3/2}$ reported in other studies \citep{haugen2004simulations,borovsky2012velocity}, possibly due to the pronounced bottleneck effect from using hyperdiffusion \citep{frisch2008hyperviscosity}. 

Without placing too much emphasis on the specific values of $\sigma_u$ and $\sigma_b$ for now, we wish to check if the scaling we derived in eq. \eqref{eq:scaling_decopled} is consistent with the $\sigma_u$ and $\sigma_b$ we observe in our simulations. Figure \ref{fig:vis_res_pr}(a),(b) (also Fig. \ref{fig:vis_res_pr-abc} in Appendix) shows the effective transport coefficients $\nu_t$, $\eta_t$, and $Pr_t$ as a function of scale calculated using their respective definitions in eqs. \eqref{eq:pi_u_nu}-\eqref{eq:pr_def}. We can see that $\nu_t(k)\sim k^{-5/3}$ (or $\sim k^{-3/2}$) and $\eta_t(k)\sim k^{-4/3}$, consistent with relation \eqref{eq:scaling_decopled} when $\sigma_u=1/6$ (or $\sigma_u=1/4$) and $\sigma_b=1/3$ as in our simulations. Moreover, we see in Figure \ref{fig:vis_res_pr}(a),(b) (and Fig. \ref{fig:vis_res_pr-abc} in Appendix) that  $Pr_t(k)\sim k^{-1/3}$ (or $\sim k^{-1/6}$), which is also consistent with the derived scaling in eq. \eqref{eq:scaling_decopled} with $\sigma_u=1/6$ (or $\sigma_u=1/4$) and $\sigma_b=1/3$ in our simulated flows. Panels (c)-(d) in Figure \ref{fig:vis_res_pr} also show $\nu_t$, $\eta_t$, and $Pr_t$ but calculated from eqs. \eqref{eq:TurbViscDef2}-\eqref{eq:TurbResistDef2}. Turbulent resistivity is very similar to that in Fig. \ref{fig:vis_res_pr}(a),(b) with a $\eta_t\sim k^{-4/3}$ scaling, whereas $\nu_t$ has a scaling that is slightly shallower than that in Fig.~\ref{fig:vis_res_pr}(a),(b). Since $Pr_t=\nu_t/\eta_t$, it is sensitive to slight changes in the scaling with $Pr_t\sim k^{-1/3}$ only over the decoupled range $k\in[50,200]$ but not for smaller $k$. 

Qualitatively, the scalings
of transport coefficients in Fig.~\ref{fig:vis_res_pr}(c),(d) are consistent with those in  Fig.~\ref{fig:vis_res_pr}(a),(b), generally increasing at larger scales. 
We believe that this agreement between the different definitions of transport coefficients will be enhanced as the dynamic range increases
and more definitive  power-law scalings emerge. Indeed, we will present evidence below that the dynamic range in simulations that are possible today, including ours here, do not yet have a converged power-law scaling.

The scaling of $\eta_t\sim k^{-4/3}$ agrees with our eq. \eqref{eq:scaling_solar} applicable to the solar wind, as does $\nu_t\sim k^{-3/2}$ from Runs $\rm II_c$ and $\rm  V_d$.
The scaling of $\nu_t\sim k^{-5/3}$ in Runs $\rm I_c$, $\rm III_c$, and $\rm IV_c$ decays faster than $k^{-3/2}$ in eq. \eqref{eq:scaling_solar} since $\sigma_u<1/4$ in those simulations, associated with a shallower spectrum. This may be attributed to the bottleneck effect from hyperviscosity \citep{frisch2008hyperviscosity}, which produces a pileup at the small scales (see the spectra in Fig. \ref{fig:spec_u} of Appendix).

\begin{figure*}[!ht]
\gridline{\fig{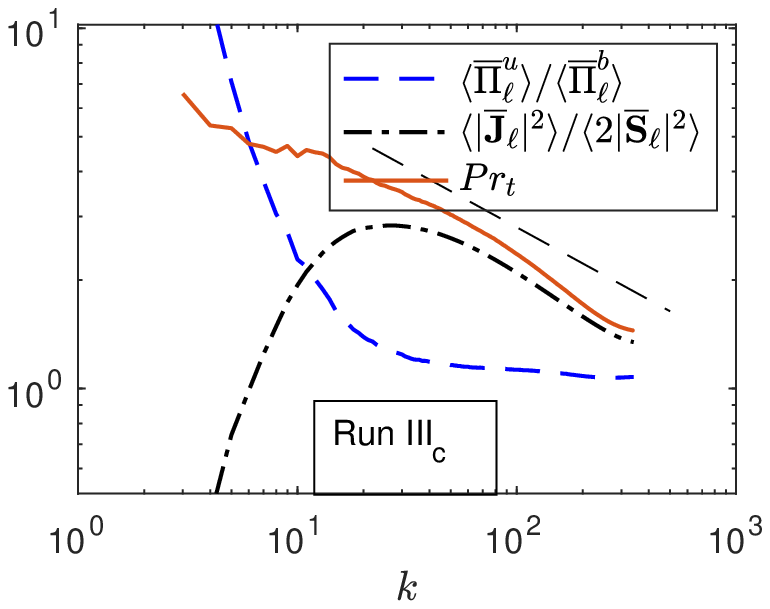}{0.35\textwidth}{}
          \fig{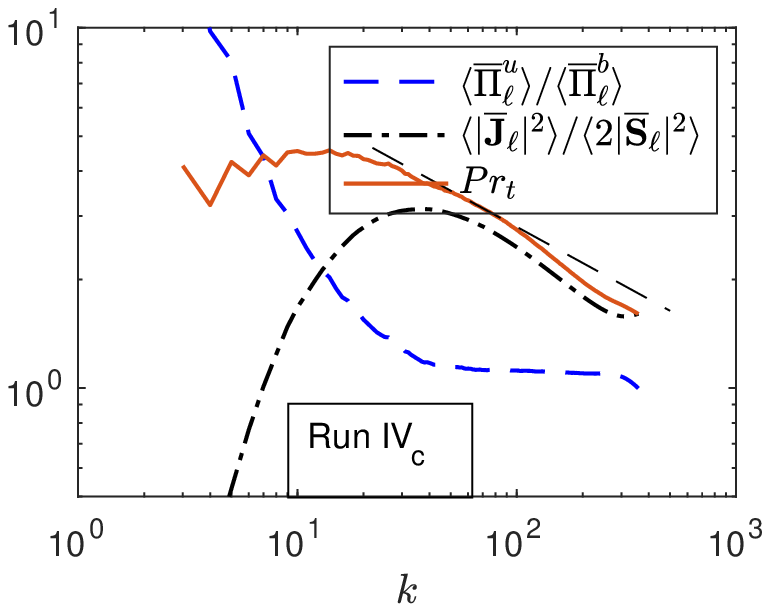}{0.35\textwidth}{}
          }
\caption{Plots showing $Pr_t$ and its two components $\langle \OL \Pi^u_\ell \rangle / \langle \OL \Pi^b_\ell \rangle$ and $\langle |\OL \bJ_\ell|^2 \rangle/ \langle 2|\OL \bS_\ell|^2\rangle$ at the highest resolutions of Run III and IV (Taylor-Green forcing) in Table \ref{Tbl:Simulations}. A reference line with a slope of -1/3 (black dashed) is added. The plots show that $\langle \OL \Pi^u_\ell \rangle / \langle \OL \Pi^b_\ell \rangle$ approaches a constant in the decoupled range. Simulations with helical forcing are shown in Fig. \ref{fig:pi_spec_pr-abc} in Appendix.
}
\label{fig:pi_spec_pr}
\end{figure*}

Figure \ref{fig:vis_res_pr}(a)(b) also shows $Pr_t$ larger than unity in the inertial-inductive range, decreasing to $Pr_t\approx1\text{~to~}2$ at the smallest inertial-inductive scales $\ell_d$ in all cases (see also Table \ref{Tbl:Pr_t_kd} in Appendix), where $\ell_d$ is defined as the scale where $ \langle\OL \Pi^u_{\ell} + \OL\Pi^b_{\ell}\rangle = \nu \langle|\nabla\OL{\textbf u}|^2\rangle + \eta \langle|\nabla\OL{\textbf B}|^2\rangle$. For non-unity $Pr_m$, $\ell_d\equiv \text{max} (\ell_{\nu}, \ell_{\eta})$. $\ell_{\nu}$ and $\ell_\eta$ are defined as scales where $ \langle\OL \Pi^u_\ell\rangle = \nu \langle|\nabla\OL{\textbf u}|^2\rangle$ and $\langle \OL\Pi^b_\ell\rangle = \eta \langle|\nabla\OL{\textbf B}|^2\rangle$.

Figure \ref{fig:pi_spec_pr} (and Fig. \ref{fig:pi_spec_pr-abc} in Appendix) shows ratios $\langle \OL \Pi^u_\ell \rangle / \langle \OL \Pi^b_\ell \rangle$ and $\langle |\OL \bJ_\ell|^2 \rangle/  \langle 2|\OL \bS_\ell|^2\rangle$, the product of which yields $Pr_t$ in eq. \eqref{eq:pr_def}. $\langle \OL \Pi^u_\ell \rangle / \langle \OL \Pi^b_\ell \rangle$ becomes constant in the decoupled range due to the conservative (constant) KE and ME cascades in this range. $\langle |\OL \bJ_\ell|^2 \rangle/ \langle 2|\OL \bS_\ell|^2\rangle$ is equal to $\int_{0}^{k} k'^2 E^{b}(k') dk'/ \int_{0}^{k} k'^2 E^{u}(k') dk'$. The ratio $\langle |\OL \bJ_\ell|^2 \rangle/ \langle 2|\OL \bS_\ell|^2\rangle$ increases because $E^{b}(k)< E^{u}(k)$ at forcing scales (forcing in velocity field) but  $E^{b}(k)$ catches up and exceeds $E^u(k)$ at larger $k$. The ratio $\langle |\OL \bJ_\ell|^2 \rangle/ \langle 2|\OL \bS_\ell|^2\rangle$ decays after reaching a peak since (1) each of $\OL \bJ_\ell$ and $\OL \bS_\ell$ is dominated by the largest wavenumbers below the cutoff $k<L/\ell$, and (2) $E^{u}(k)$ is shallower than $E^{b}(k)$ at high $k$ in the inertial-inductive range, making $|\OL \bS_\ell|^2$ grow faster than $|\OL \bJ_\ell|^2$ as $\ell\to0$.

\section{Discussion}\label{sec:discussion}
We now provide the physical explanation for why $Pr_t$ seems to increase at larger scales and discuss whether or not this trend is expected to persist for an arbitrarily wide dynamical range ($Re\to\infty$). As we have mentioned above, $\sigma^{u}$ and $\sigma^{b}$ are a measure of the velocity and magnetic fields' smoothness, respectively \citep{Aluie17}.
If $\sigma_u < \sigma_b$ (corresponding to a shallower scaling of $E^u(k)$ relative to $E^b(k)$) as in our simulations and many other independent reports from solar wind observations and simulations \citep[e.g.,][]{podesta2007spectral,mininni2009finite,borovsky2012velocity,grappin2016alfven}, then the velocity field is rougher than the magnetic field.
This implies that small-scale velocity ``eddies'' have a higher \emph{proportion} of the overall kinetic energy compared to the \emph{proportion} small-scale magnetic ``eddies'' contribute to the overall magnetic energy (i.e., $\frac{\text{small-scale\ KE}}{\text{total\ KE}} > \frac{\text{small-scale\ ME}}{\text{total\ ME}}$). Note that the latter statement is not based on comparing $E^u(k)$ to $E^b(k)$ at high $k$ in absolute terms, where we see $E^b(k) \gtrsim E^u(k)$. Rather, it is based on the strength of ``eddies'' \emph{relative} to the overall velocity or magnetic field, respectively. 
 
The coarse-grained strain and current, $\OL \bS_\ell$ and $\OL \bJ_\ell$, are cumulative quantities, i.e., they include the contribution from \emph{all} scales larger than $\ell$, for any $\ell$. It follows from the above paragraph that as the coarse-graining $\ell$ is made smaller, the relative contribution from scales near $\ell$ to $ |\OL \bS_\ell|^2$ is more significant than to $ |\OL \bJ_\ell|^2$.
From the definition of $Pr_t$ in eq. \eqref{eq:pr_def} and with $\langle \OL\Pi^u_\ell\rangle/\langle \OL\Pi^b_\ell\rangle$ being scale-independent in the decoupled range, we have $Pr_t \propto \langle |\OL \bJ_\ell|^2 \rangle/  \langle 2 |\OL \bS_\ell|^2\rangle$ in the decoupled range. Clear evidence of this is shown in Fig. \ref{fig:pi_spec_pr} (and Figs. \ref{fig:Pr_effects},\ref{fig:pi_spec_pr-abc}). As $\ell$ decreases (or $k$ increases), both $ |\OL \bS_\ell|^2$ and $ |\OL \bJ_\ell|^2$ increase because contributions from $<\ell$ are included. However, due to larger roughness of the velocity field, the increase in $|\OL \bS_\ell|^2$ is more pronounced than that in $|\OL \bJ_\ell|^2$, leading to a decrease in the ratio $\langle |\OL \bJ_\ell|^2 \rangle/  \langle 2 |\OL \bS_\ell|^2\rangle$. This explains why $Pr_t$ seems to decrease with larger $k$ over the decoupled range (range over which each of $\langle \OL\Pi^u_\ell\rangle$ and $\langle \OL\Pi^b_\ell\rangle$ is scale-independent).
 
In the conversion range over which $\langle \OL\Pi^u_\ell\rangle$ and $\langle \OL\Pi^b_\ell\rangle$ are still varying with $\ell$, the scaling of $Pr_t$ depends on both $ \langle\OL \Pi^u_{\ell}\rangle / \langle\OL \Pi^b_{\ell}\rangle$ and $\langle |\OL \bJ_\ell|^2 \rangle/  \langle 2|\OL \bS_\ell|^2\rangle$. On the one hand, $\langle\OL \Pi^u_{\ell}\rangle/ \langle\OL \Pi^b_{\ell}\rangle > 1$ since energy is input into the velocity field at the largest scales and more kinetic energy is cascading compared to magnetic energy, such that $\langle\OL \Pi^u_{\ell}\rangle/ \langle\OL \Pi^b_{\ell}\rangle \to \infty$ as $\ell \to \ell_f$ approaching the forcing scale $\ell_f$. On the other hand, we have the ratio $\langle |\OL \bJ_\ell|^2 \rangle/  \langle 2|\OL \bS_\ell|^2\rangle$ decreasing in that limit of $\ell \to \ell_f$ since the strain becomes relatively stronger at the forced scales. From Fig. \ref{fig:pi_spec_pr}, we find that in our simulated flows, the $Pr_t$ scaling over the conversion range either decaying weakly or flat as $k$ increases. Since the conversion range is limited in extent and does not increase with an increasing dynamic range \citep{bian2019decoupled}, it is not very meaningful to discuss a scaling of $Pr_t$ over this range.

Crude estimates of the competition between large scale magnetic flux advection and large scale magnetic flux diffusion in accretion disks require
$Pr_t (R/H)>1$ (where $H$ is the disk scale height and $R$ is the disk radius) for the former to be competitive with the latter in the disk interior
\citep{Lubow+1994,Blackman+2015}.\footnote{The turbulent Prandtl number used in \cite{Lubow+1994} is the inverse of $Pr_t$}
That we find values of $Pr_t >1$   
means that  large-scale MHD flow may be  more efficient at advecting large scale magnetic flux while shedding angular momentum outward (via $\nu_t$)
than would be the case for  $Pr_t \le 1$.
That said, pinning down the exact implications are difficult given the additional dependence of disk physics on stratification with the possibility of flux advection in surface layers \citep[e.g.,][] {Lovelace+2009,Zhu+2018}.

\subsection{$Pr_t$ scaling under different flow conditions}

We have tested the scaling of $Pr_t$ under different microscopic $Pr_m$ flow conditions. We remind the reader that our results here pertain to the decoupled range, which is within the inertial-inductive range. These scales are immune from the direct influence of both resistivity and viscosity. We do not expect our results here (and those of \cite{bian2019decoupled} upon which this analysis is based) to hold in the viscous-inductive (Batchelor) range at high $Pr_m$, or in the inertial-resistive range at low $Pr_m$.
From practical modeling considerations, such as when simulating a galactic accretion disc at global scales, grid-resolution constraints render it virtually impossible to have $\Delta x\sim \ell$ within the viscous-inductive range. Therefore, the restriction of our analysis to inertial-inductive scales is still pertinent to modeling as well as being of theoretically import. 

\begin{figure*}[!ht]
\gridline{\fig{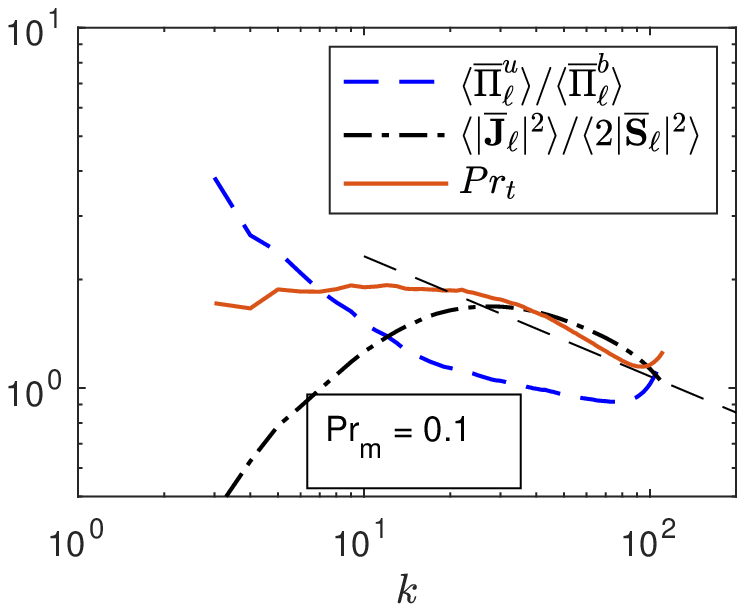}{0.35\textwidth}{}
          \fig{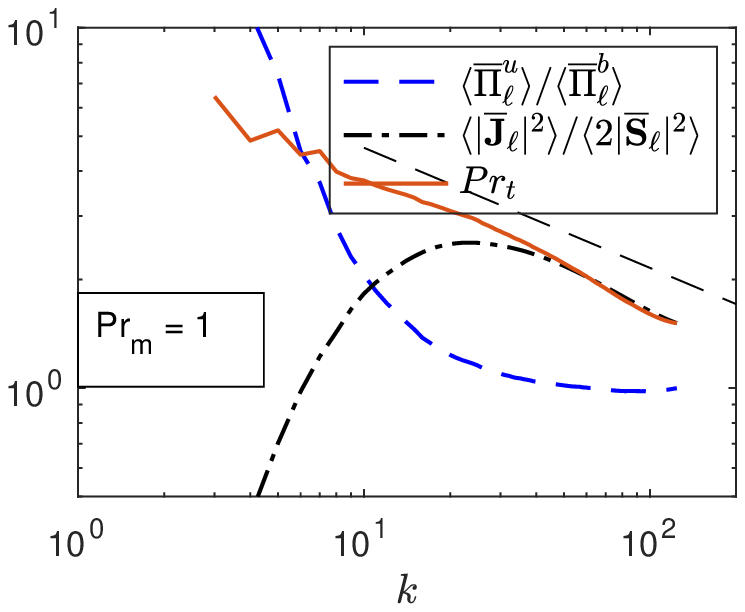}{0.35\textwidth}{}
          }
\gridline{\fig{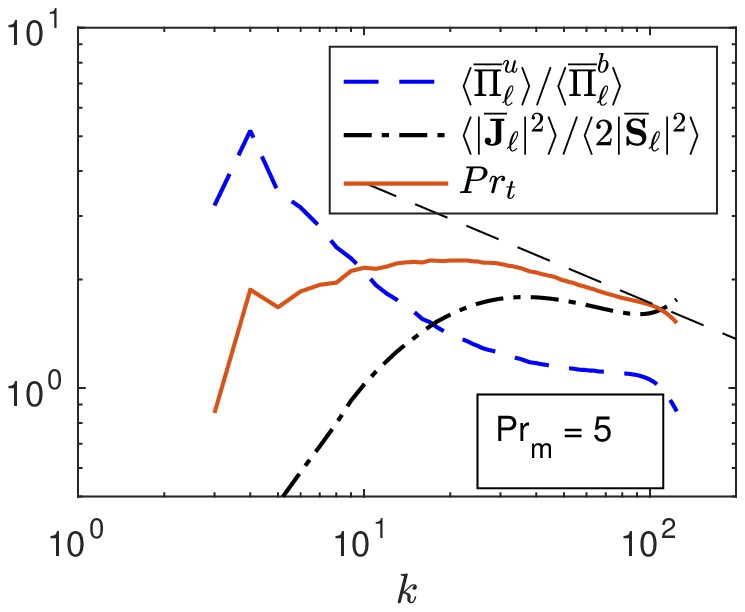}{0.35\textwidth}{}
          \fig{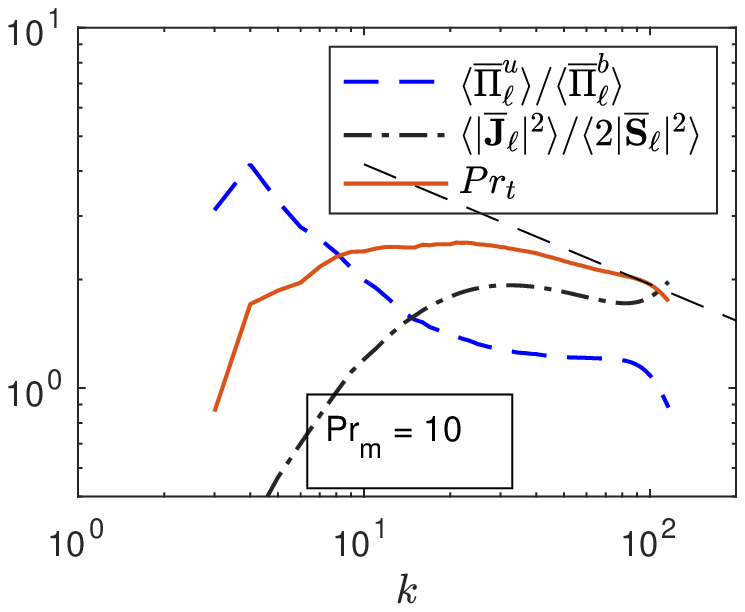}{0.35\textwidth}{}
          }
\caption{Plots showing $Pr_t$ and its two components $\langle \OL \Pi^u_\ell \rangle / \langle \OL \Pi^b_\ell \rangle$ and $\langle |\OL \bJ_\ell|^2 \rangle/ \langle 2|\OL \bS_\ell|^2\rangle$ using unity ($Pr_m$ = 1)  and non-unity microscopic Prandtl numbers ($Pr_m$ = 0.1, 5, 10) on $512^3$ grid. See parameters in Table \ref{Tbl:Simulations}. A reference line with a slope of -1/3 (black dashed line) is added.}
\label{fig:Pr_effects}
\end{figure*}

Fig. \ref{fig:vis_res_pr}(a)(b) shows the case (Run IV) of $Pr_m=2$, where we find a scaling of $Pr_t$ similar to the case of unity microscopic $Pr_m$. We also conduct simulations (Table \ref{Tbl:Simulations}) at 
$Pr_m$ = 0.1, 5, and 10, albeit at a lower resolution of $512^3$ due to the computational overhead required by non-unity $Pr_m$. Fig. \ref{fig:Pr_effects} shows that scale-dependence of $Pr_t$ is consistent with that of unity-$Pr_m$ runs, although the scaling is not as clear. Due to the lower resolution, the decoupled range is barely established in the non-unity $Pr_m$ cases. Non-unity $Pr_m$ simulations require even larger Reynolds numbers to achieve a significant decoupled range and still make an allowance for a viscous-inductive or an inertial-resistive range of scales. This is beyond our computing capability for this work.

Our results also suggest that within the limited dynamic range of our simulations, increasing the external B-field strength from 0 (Run I) to 2 (Run V) to 10 (Run II) seems to change the $Pr_t$ scaling slightly from $k^{-1/3}$ to $k^{-1/6}$ due to $\sigma_u$ increasing from 1/6 to 1/4 (see Fig. \ref{fig:vis_res_pr-abc} in Appendix). {However, we do not believe this trend will persist at asymptotically high-$Re$ as we discuss in the following subsection.}
We also note that our analysis here does not distinguish the anisotropy in turbulent transport. Our effective transport coefficients in this paper are isotropic even though the underlying turbulent flow may be anisotropic such as in Runs II and V (see Fig. \ref{fig:vis} in Appendix). We hope this work is extended to anisotropic turbulent transport in future studies. 

\subsection{$Pr_t$ scaling at asymptotically high-$Re$}
Can we expect the scaling of $Pr_t$ in Fig. \ref{fig:vis_res_pr}(a),(b), which is in support of our relations in eq. \eqref{eq:scaling_decopled}, to  extrapolate to the wide dynamical ranges (high-$Re$) that exist in many astrophysical systems of interest?

Figure \ref{fig:Pr_diffRes_unnorm} {(and Fig. \ref{fig:Pr_diffRes_unnorm-abc} in Appendix)} examines the scaling of $Pr_t(k)$ at different Reynolds numbers. Each panel shows results from a suite of simulations under the same parameters except for $Re$ (or grid-resolution). The plots show that $Pr_t(k)$ takes on a value between 1 and 2 at the smallest scales within the inertial-inductive range, regardless of $Re$ (also Fig. \ref{fig:Pr_diffRes_norm} and Table \ref{Tbl:Pr_t_kd} in Appendix). These scales near $\ell_d$ are bordering the dissipation range. The reason $Pr_t(k\approx L/\ell_d)\approx 1\text{~to~}2$ can be understood from definition \eqref{eq:pr_def} of $Pr_t = \left(\langle\OL \Pi^u_{\ell}\rangle / \langle\OL \Pi^b_{\ell}\rangle\right) \left(\langle |\OL \bJ_\ell|^2 \rangle/  \langle 2|\OL \bS_\ell|^2\rangle\right)$. Due to equipartition of the cascades in the decoupled range, we have $\langle\OL \Pi^u_{\ell}\rangle / \langle\OL \Pi^b_{\ell}\rangle \approx 1$, whereas $\langle |\OL \bJ_\ell|^2 \rangle/  \langle 2|\OL \bS_\ell|^2\rangle \approx 1\text{~to~}2$, as is clear from Fig. \ref{fig:pi_spec_pr} (and Figs. \ref{fig:Pr_effects},\ref{fig:pi_spec_pr-abc}). The latter 
can also be inferred from comparing the spectra $E^{u}(k')$ and $E^{b}(k')$ in Figs. \ref{fig:spec_u}-\ref{fig:spec_b} via eqs. \eqref{eq:nu_spec_u}-\eqref{eq:eta_spec_b}.

\begin{figure*}[!ht]
\gridline{\fig{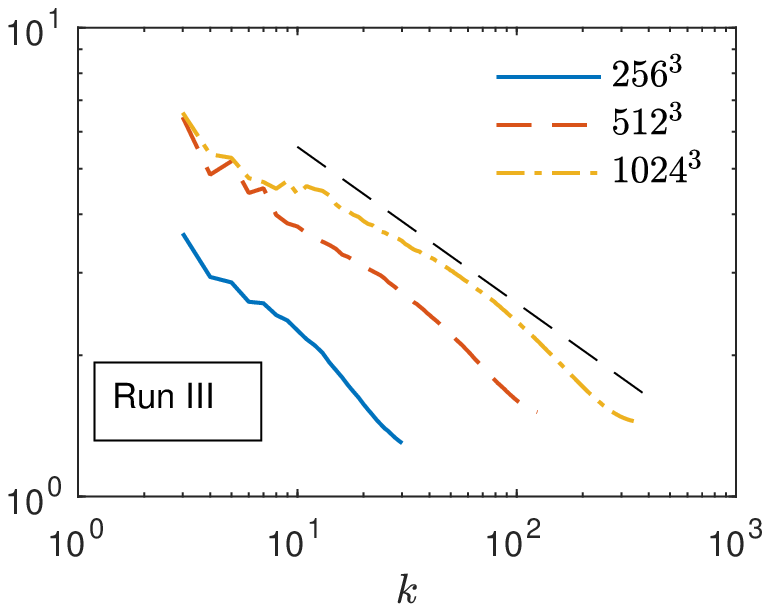}{0.35\textwidth}{}
          \fig{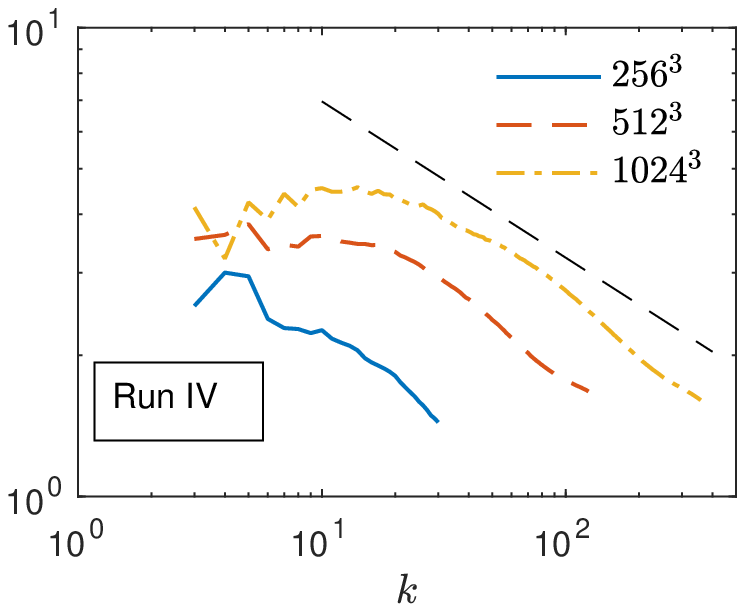}{0.35\textwidth}{}
          }
\caption{Plots showing $Pr_t$ at different Reynolds numbers (grid resolution) of Run III and IV (Taylor-Green forcing) in Table \ref{Tbl:Simulations}. A reference line with a slope of -1/3 (black dashed) is added. 
Simulations with helical forcing are shown in Fig. \ref{fig:Pr_diffRes_unnorm-abc} in Appendix.}
\label{fig:Pr_diffRes_unnorm}
\end{figure*}

Based on these observations, it is physically reasonable to assume that for a sufficiently wide dynamical range (or large $Re$), $\langle |\OL \bJ_{\ell_d}|^2 \rangle/  \langle 2|\OL \bS_{\ell_d}|^2\rangle$ converges to a constant when $\ell\approx \ell_d$ (or $k\approx k_d$), independent of the dynamical range extent (i.e., independent of $Re$). 
That is, the ratio $\langle |\OL \bJ_{\ell_d}|^2 \rangle/  \langle 2|\OL \bS_{\ell_d}|^2\rangle$ has the same value under successive refinement of the grid. This effectively provides us with a conceptual boundary condition on $Pr_t(k)$ at those smallest scales $k\approx k_d$. 

These logical considerations, when combined with the scaling $Pr_t(k)\sim k^{-2(\sigma_b-\sigma_u)}$ in eq. \eqref{eq:scaling_decopled} (with empirical support in Figs. \ref{fig:vis_res_pr}(a)(b),\ref{fig:vis_res_pr-abc}), may lead us at face value to the uncomfortable conclusion that at any fixed $k$ in the inertial-inductive range, $Pr_t(k)$ will keep increasing with increasing $Re$ (or higher resolution) as Fig. \ref{fig:Pr_diffRes_unnorm} (and Fig. \ref{fig:Pr_diffRes_unnorm-abc} in Appendix) seems to indicate. That is unless the MHD dynamics eventually yields $\sigma_u=\sigma_b$ in this asymptotic limit, i.e., at successively higher resolution simulations. {Indeed, there are indications from Figs. \ref{fig:Pr_diffRes_unnorm},\ref{fig:Pr_diffRes_unnorm-abc} that the $Pr_t(k)$ sensitivity to $Re$ decreases with increasing $Re$ as we now discuss.}

At first glance, Fig. \ref{fig:Pr_diffRes_unnorm} seems to indicate that 
$Pr_t(k)$ at any fixed wavenumber, e.g., $k=50$ within the decoupled range, $Pr_t(k=50)$ increases with increasing resolution. Yet, as we shall now argue, Fig. \ref{fig:Pr_diffRes_unnorm} highlights how certain metrics such as $Pr_t$ in our simulations, which are very high-resolution by today's standards, are still not fully converged to the high-$Re$ limit. From the definition of $Pr_t$ in eq. \eqref{eq:pr_def}, this increase can only be due to an increase of the cascade ratios $\langle\OL \Pi^u_{\ell}\rangle / \langle\OL \Pi^b_{\ell}\rangle$, or the current-to-strain ratio, 
$\langle |\OL \bJ_\ell|^2 \rangle/  \langle 2|\OL \bS_\ell|^2\rangle$, or both. We find in Fig. \ref{fig:pi_js_ratio} that the latter accounts for most of this increase. Fig. \ref{fig:pi_js_ratio} suggests that the cascade ratio $\langle\OL \Pi^u_{\ell}\rangle / \langle\OL \Pi^b_{\ell}\rangle$ is fairly converged with resolution in our largest simulations at $\ell=L/50$. Physically, we expect $\langle |\OL \bJ_\ell|^2 \rangle/  \langle 2|\OL \bS_\ell|^2\rangle$ to also converge since the ratio depends on the strain and current (or equivalently, the spectra) at scales larger than $L/50$. These should not remain sensitive to the smallest scales in a simulation once a sufficiently high resolution has been achieved. Fig. \ref{fig:pi_js_ratio} in Appendix indicates that the high resolution of our simulations is still not sufficient for the convergence of these quantities ($\langle 2|\OL \bS_\ell|^2\rangle$ and $\langle |\OL\bJ_\ell|^2\rangle$).  Ignoring convergence trends under the guise of ``having conducted the highest resolution simulation to date'' can be rife with pitfalls. In general, when analyzing simulations of turbulent flows, it is vitally important to study trends as a function of Reynolds number and check if the phenomenon under study persists and can be extrapolated to the large Reynolds numbers present in nature.

What conclusion on $Pr_t(k)$ scaling do these convergence considerations lead us to? If we accept that with increasing resolution, $Pr_t(k_*)$ has to converge to a specific value for any fixed $k_*$ within the inertial-inductive range, and if we also accept that at the smallest scales within the inertial-inductive range $\approx\ell_d$, $Pr_t(k_d)$ also converges to a constant value, then as the gap between $k_*$ and $k_d$ widens with a wider dynamical range ($k_d\to\infty$), we must have that $Pr_t(k)\sim k^{-2(\sigma_b-\sigma_u)}\sim k^0$ approach a $k$-independent scaling with $\sigma_b=\sigma_u$ in the asymptotic limit $Re\to\infty$. {Our Figs. \ref{fig:Pr_diffRes_unnorm},\ref{fig:Pr_diffRes_unnorm-abc} lend some support to our assertion as they show that the $Pr_t(k)$ is converg\emph{ing} (but not converg\emph{ed}) at the largest scales with increasing resolution.}

Such a conclusion would have wide-ranging implications, foremost of all regarding the power-law scaling of spectra in MHD turbulence. However, it is important that our results are further verified by the community under different parameter conditions, e.g., $B_0$ strength and $Pr_m$, and perhaps also from higher resolution simulations.

\section{Conclusion}\label{sec:conclusion}

In this paper, we are proposing a somewhat new method to measure turbulent transport coefficients (turbulent viscosity $\nu_t$, resistivity $\eta_t$, and magnetic Prandtl number $Pr_t$) at different scales using the coarse-graining approach. To our knowledge, this is the first determination of $Pr_t$ as a function of length scale. From analyzing the kinetic and magnetic energy cascade rates, we infer power-law scaling in eq. \eqref{eq:scaling_decopled} for $\nu_t$, $\eta_t$, and $Pr_t$ given our definitions of those transport coefficients.  This approach circumvents relying on particular values for the spectral scaling exponents ($\sigma_{u}$ and $\sigma_{b}$) from a specific MHD phenomenology --whether it exists or not-- by relying on results from \cite{bian2019decoupled} of conservative KE and ME cascades. Our analysis here relied on high-resolution DNS under different forcing, external B-field strength, and microphysical $Pr_m$.

Our DNS results indicate that $Pr_t\approx1 \text{~to~}2$ at the smallest inertial-inductive scales, increasing to $Pr_t\approx5 \text{~to~}10$ at the largest scales. For accretion disks, conservative minimalist estimates for advection of large scale vertical magnetic fields to win over turbulent diffusion require  $Pr_t (H/R)>1$, so that  larger values of  $Pr_t$
improve the efficacy of flux advection over diffusion \citep[e.g.,][]{Lubow+1994}. This condition and the direct applicability of our specific results are both textured by
detailed disk physics \citep[e.g.,][]{Zhu+2018}, 
including stratification, not studied here.

Nevertheless, based on physical considerations, our analysis suggests that $Pr_t$ has to become scale-independent and of order unity in the decoupled range at sufficiently high Reynolds numbers (or grid-resolution), and that the power-law scaling exponents of velocity and magnetic spectra become equal.

If indeed the power-law scaling exponents of velocity and magnetic spectra ($\sigma_u$ and $\sigma_b$) become equal in the $Re\to \infty$  limit, it would have wide-ranging implications, foremost of all regarding the power-law scaling of spectra in MHD turbulence \citep{PolitanoPouquet98a,PolitanoPouquet98b,Aluie17}. However, as discussed above, our $Pr_t$ scaling is not quite converged, despite showing a converging trend. It is important for our results to be further checked by the community using simulations of higher resolution and for a wider range of parameters, e.g., $\bB_0$ strengths and $Pr_m$ values.

Our results also suggest that the presence of a mean B-field does not affect $Pr_t$ significantly. However, we only consider $Pr_t$ as a scalar in this study. \cite{LesurLongaretti09} considered an anisotropic turbulent resistivity tensor with an external B-field. Under non-unity microphysical $Pr_m$, our results are consistent with those of $Pr_m=1$, although we could not establish a clear decoupled range due to insufficient simulation resolution.

In addition to potential implications for astrophysical systems, our analysis of how $\nu_t$, $\eta_t$, and $\Pr_t$ vary with length-scale provides a practical model for these quantities that does not rely on any particular MHD turbulence phenomenology.

The simulations we conducted here are fairly idealized (incompressible flows in a periodic domain with artificial forcing). We hope that this work offers a path to analyzing more complicated flows since our method can be applied to more realistic simulations such as of global accretion disk flows.  For the pursuit of isotropic diffusion coefficients,  measuring $\nu_t$ and $\eta_t$ at any length-scale from eqs. \eqref{eq:pi_u_nu},\eqref{eq:pi_b_eta}  does not require the existence of an inertial range or even turbulence, even though in the present paper we applied the method to a case of fully developed turbulence. For some applications, we believe that our approach  complements existing approaches such as test-field methods \citep{schrinner2007mean,kapyla2020turbulent}  of measuring  turbulent transport. These methods  involve taking the velocities computed from a numerical simulation and then separately solving for the transport coefficients using an imposed test magnetic field. Traditionally these have been  restricted to the kinematic regime of a weak magnetic field (although see \cite{Kapyla+2021}).

Finally, our work should not be construed as an endorsement of the ``eddy viscosity/resistivity'' model wherein turbulent processes $\OL\btau_\ell$ and $\OL{\bepsilon}_\ell$ representing scales $<\ell$ are modeled as purely diffusive. Our approach can be extended to models in which transport is not entirely diffusive, such as those which include the the helical $\alpha$-effect.

\acknowledgments
This research was funded by DOE FES grants DE-SC0014318 and DE-SC0020229. 
Partial funding for this research was provided 
the Center for Matter at Atomic Pressures (CMAP), a National Science Foundation (NSF) Physics Frontier Center, under Award PHY-2020249. 
HA was also supported by NASA grant 80NSSC18K0772, DOE grant DE-SC0019329, and NNSA grants DE-NA0003914 and DE-NA0003856. 
JS was also supported by DOE grant DE-SC0019329 and NNSA grant DE-NA0003914.
EB was also supported by DOE grants DE-SC0001063, DE-SC0020432 and DE-SC0020103, and NSF grant AST-1813298.
Computing time was provided by the National Energy Research Scientific Computing Center (NERSC) under Contract No. DE-AC02-05CH11231, and by an award from the INCITE program, using resources of the Argonne Leadership Computing Facility, which is a DOE Office of Science User Facility supported under Contract DE-AC02-06CH11357.

\bibliography{main.bib}
\bibliographystyle{aasjournal}

\newpage
\appendix

This Appendix provides more details about the numerical setup, evidence of convergence, and the effects of a non-unity microscopic Prandtl number.

\section{Numerical setup}

Our numerical simulations of mechanically forced turbulence are conducted in a periodic box $\mathbb{T}^3=[0, L)^3$, with $L=2\pi$. We use a pseudo-spectral code with phase-shift dealiasing. The time integration method is a second-order Adam-Bashforth scheme. We solve the incompressible MHD equations with hyperviscosity \citep{BorueOrszag95} and hyperresistivity with a Laplacian of exponent $\alpha=5$:
\begin{gather}
\partial _t \textbf u+ ( \textbf u\bdot \grad) \textbf u =- \grad p+ \bJ \btimes \bB -\nu_h(-\nabla^2)^{\alpha} \bu + \bff,\\
\partial _t \textbf B =\grad\btimes(\textbf u\btimes \textbf B)-\eta_h(-\nabla^2)^{\alpha} \bB, \\
\grad\bdot\bu=\grad\bdot\bB=0,
\end{gather}
where $\nu_h$ is hyperviscosity, and $\eta_h$ is hyperresistivity coefficients. Hyperdiffusivity is commonly used in MHD turbulence studies \citep{ChoVishniac00,Kawai13,Beresnyak15,Meyrandetal16,kawazura2019thermal} to reduce the dissipation range extent, thereby allowing for a longer inertial-inductive range of scales. The velocity and magnetic field are initialized in $k$-space with $E^{u,b}\sim {|\bf k|^2e^{-|\bf k|^2/11}}$ spectra and random phases.

Runs I, II, and V (see Table \ref{Tbl:DetailedParameters} for simulation details) are driven by ABC forcing (named after Arnold, Beltrami, and Childress):
\begin{equation}
\textbf f \equiv [A\sin(k_f z)+C\cos(k_f y) ]\textbf e_x +[B\sin(k_f x) + A\cos(k_f z) ]\textbf e_y + [C\sin(k_f y)+B\cos(k_f x) ]\textbf e_z,
\end{equation}
where $A=B=C=0.25$, $k_f$ is forcing wavenumber, $\textbf e_x$, $\textbf e_y$, and $\textbf e_z$ are unit vectors in $x$, $y$, and $z$, respectively. ABC forcing is helical, which injects kinetic helicity into the flow. Kinetic helicity is an example of a pseudoscalar which facilitates large-scale dynamos  
\citep[e.g.][]{Parker1955,Moffatt+1978,mininni2005low,blackman2016magnetic}.

Taylor-Green (TG) forcing, which is non-helical, is used to drive the flow in Runs III and IV:
\begin{equation}
\textbf f \equiv f_0[\sin(k_f x)\cos(k_f y)\cos(k_f z) \textbf e_x -\cos(k_f x) \sin(k_f y)\cos(k_f z)\textbf e_y],
\end{equation}
where the force amplitude $f_0=0.25$. TG forcing injects no global integrated kinetic helicity into the flow.

The simulations are conducted at different Reynolds numbers with different grid resolutions. Detailed parameters are shown in Table \ref{Tbl:DetailedParameters}, where subscripts a, b, c, and d (e.g., Run $\rm V_a$ vs. $\rm V_b$ vs. $\rm V_c$ vs $\rm V_d$) denote simulations using the same parameters but at different grid resolutions and Reynolds numbers. Run I-IV are conducted with grid resolution of $256^3$, $512^3$, and $1024^3$. Run $\rm V$ is also conducted at $2048^3$ resolution. For Run III, $Pr_m=0.1$, $Pr_m=5$, and $Pr_m=10$ at grid resolution of $256^3$ and $512^3$ are added to study the effects of non-unity microscopic Prandtl number.

\begin{figure*}
\gridline{\fig{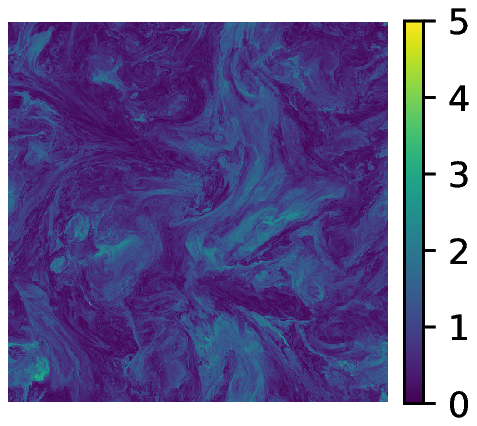}{0.3\textwidth}{(a) $|\textbf u|$ (Run $\rm I_c$)}
          \fig{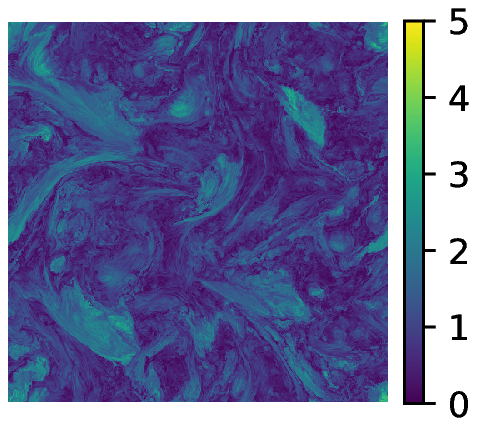}{0.3\textwidth}{(b) $|\textbf B|$ (Run $\rm I_c$)}
          }

\gridline{\fig{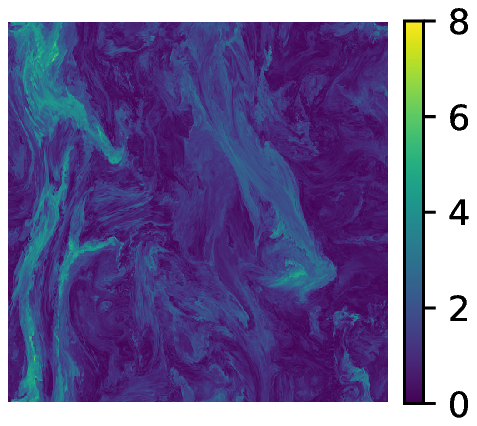}{0.3\textwidth}{(c) $|\textbf u|$ (Run $\rm II_c$)}
          \fig{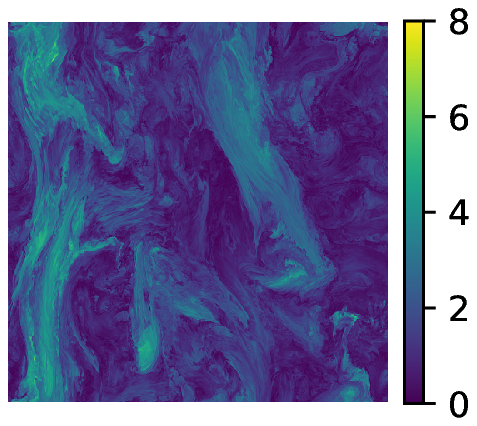}{0.3\textwidth}{(d) $|\textbf B|$ (Run $\rm II_c$)}
          }

\caption{Slices of magnitude of velocity field $|\textbf u|$ and magnetic field $|\textbf B|$. Panels (a) and (b) show results from Run $\rm I_c$ without an external B-field $|\bB_0|=0$. Panels (c) and (d) show results from Run $\rm II_c$ with $|\bB_0|=10$. The plots show significant anisotropic structures in Run $\rm II_c$.}
\label{fig:vis}
\end{figure*}

Figure \ref{fig:vis} visualizes the magnitude of velocity and magnetic fields ($|\textbf u|$ and $|\textbf B|$) in two simulations. The anisotropic structures are significant with the presence of an external magnetic field (Fig. \ref{fig:vis}(c,d)).

\begin{figure*}
\gridline{\fig{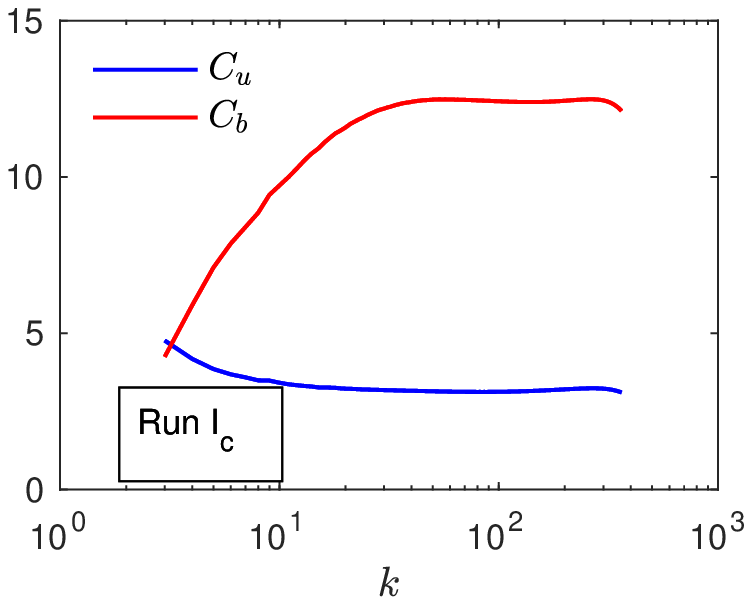}{0.3\textwidth}{}
          \fig{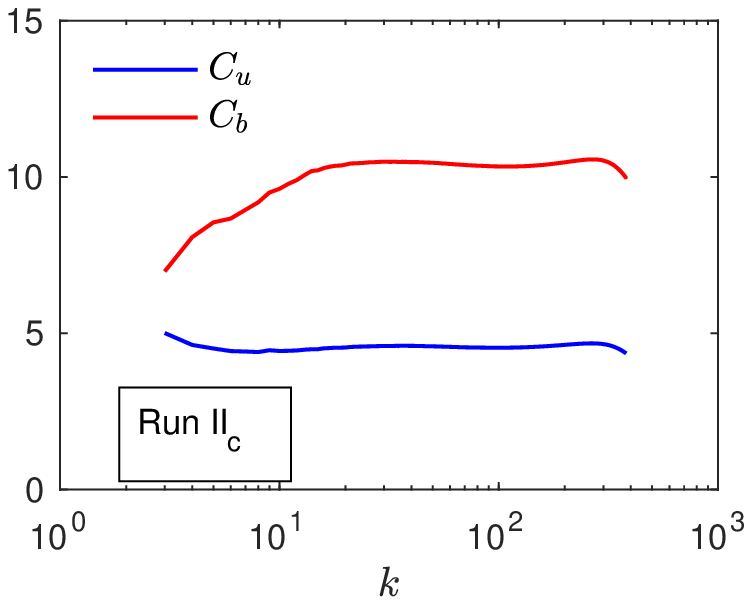}{0.3\textwidth}{}
          }
\gridline{ \fig{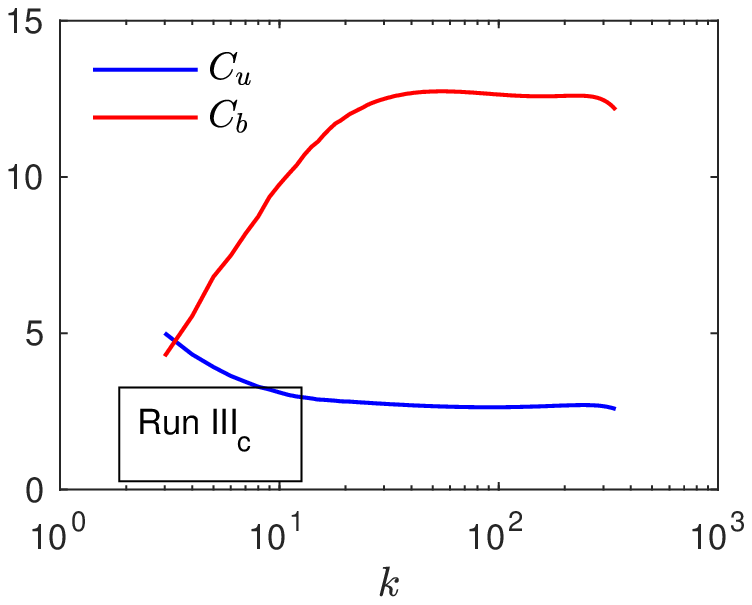}{0.3\textwidth}{}
 \fig{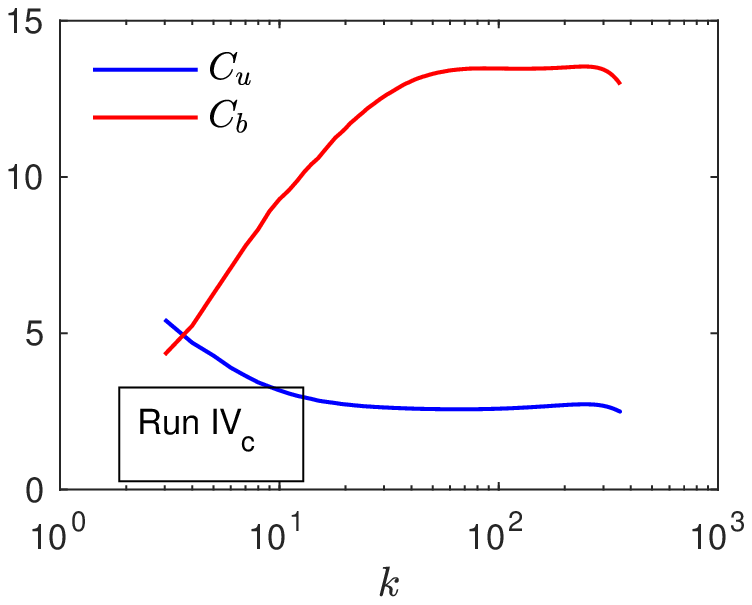}{0.3\textwidth}{}
          \fig{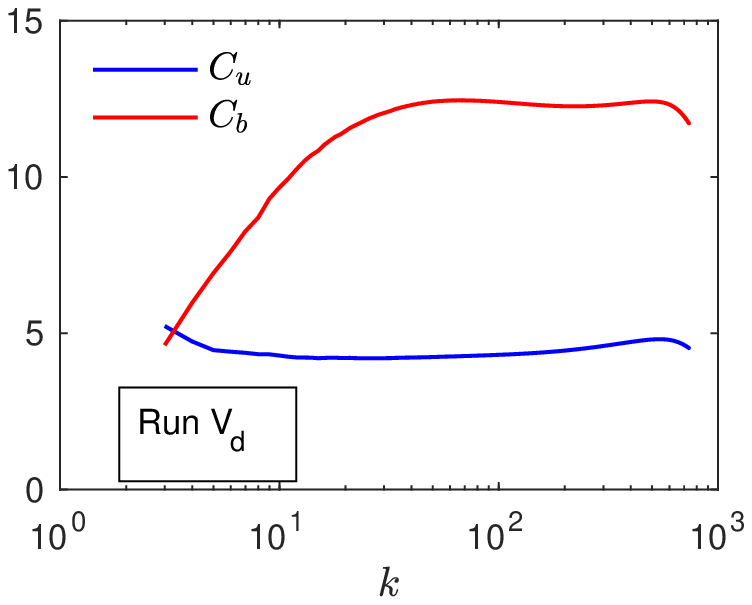}{0.3\textwidth}{}
          }
\caption{Plots showing $C_u$ and $C_b$ of Run I-V at highest resolution. $C_b$ is calculated with $\sigma_b=1/3$. $C_u$ is calculated with $\sigma_u=1/4$ in Run II$_c$ and Run V$_d$ and $\sigma_u=1/6$ in other cases.}
\label{fig:cucb}
\end{figure*}

Fig. \ref{fig:cucb} shows that $C_u$ and $C_b$ used in eqs. \eqref{eq:les_nu},\eqref{eq:les_mu} are indeed  proportionality constants that are scale-independent within the decoupled range.

\begin{table*}[!ht]
\centering
\caption{Simulations parameters: $Pr_m$ is the magnetic Prandtl number. $B^\smax_k=\sqrt{\max_k[E^b(k)]}$ is at the magnetic spectrum's [$E^b(k)$] peak. ABC (helical) and TG (non-helical) forcing are applied at wavenumber $k_f$.}
\begin{tabular}{lcccccccc}
\hline
\hline
Run  & Grid  & Forcing                    & $k_f$ & $Pr_m$ & $|\bB_0|/B^\smax_k$        & $\nu_h$   & $\eta_h$ \\ \hline
   $\rm I_a$     & $256^3$  & ABC           & 2    & 1      & 0   & $5\times 10^{-16}$  & $5\times 10^{-16}$ \\
   $\rm I_b$     & $512^3$  & ABC           & 2    & 1      & 0   & $2\times 10^{-21}$  & $2\times 10^{-21}$  \\
   $\rm I_c$     & $1{,}024^3$  & ABC       & 2    & 1      & 0   & $4\times 10^{-25}$  & $4\times 10^{-25}$   \\
   $\rm II_a$    & $256^3$  & ABC           & 2    & 1      & 10  & $5\times 10^{-16}$  & $5\times 10^{-16}$ \\
   $\rm II_b$   & $512^3$  & ABC            & 2    & 1      & 10  & $2\times 10^{-21}$  & $2\times 10^{-21}$   \\
   $\rm II_c$   & $1{,}024^3$  & ABC        & 2    & 1      & 10  & $4\times 10^{-25}$  & $4\times 10^{-25}$ \\
   $\rm III_a$   & $256^3$  & TG            & 1    & 1      & 0   & $5\times 10^{-16}$  & $5\times 10^{-16}$  \\
   $\rm III_b$  & $512^3$  & TG             & 1    & 1      & 0   & $2\times 10^{-21}$  & $2\times 10^{-21}$    \\
   $\rm III_c$  & $1{,}024^3$  & TG         & 1    & 1      & 0   & $4\times 10^{-25}$  & $4\times 10^{-25}$    \\
   $\rm IV_a$    & $256^3$  & TG           & 1    & 2      & 0   & $2\times 10^{-16}$  & $1\times 10^{-16}$ \\ 
   $\rm IV_b$    & $512^3$  & TG           & 1    & 2      & 0   & $4\times 10^{-21}$  & $2\times 10^{-21}$   \\ 
   $\rm IV_c$    & $1{,}024^3$  & TG       & 1    & 2      & 0   & $4\times 10^{-25}$  & $2\times 10^{-25}$  \\
   $\rm IV_a$($Pr_m$ = 0.1)   & $256^3$  & TG            & 1    & 0.1    & 0   & $2\times 10^{-17}$   & $2\times 10^{-16}$  \\
   $\rm IV_b$($Pr_m$ = 0.1)   & $512^3$  & TG            & 1    & 0.1    & 0   & $2\times 10^{-21}$  & $2\times 10^{-20}$    \\
   $\rm IV_a$($Pr_m$ = 5)   & $256^3$  & TG            & 1    & 5      & 0   & $1\times 10^{-16}$  & $2\times 10^{-17}$  \\
   $\rm IV_b$($Pr_m$ = 5)   & $512^3$  & TG            & 1    & 5      & 0   & $1\times 10^{-20}$  & $2\times 10^{-21}$    \\
   $\rm IV_a$($Pr_m$ = 10)   & $256^3$  & TG            & 1    & 10     & 0   & $2\times 10^{-16}$  & $2\times 10^{-17}$  \\
   $\rm IV_b$($Pr_m$ = 10)   & $512^3$  & TG            & 1    & 10     & 0   & $2\times 10^{-20}$  & $2\times 10^{-21}$    \\
   $\rm V_a$     & $256^3$  & ABC           & 2    & 1      & 2   & $5\times 10^{-16}$  & $5\times 10^{-16}$ \\
   $\rm V_b$     & $512^3$  & ABC           & 2    & 1      & 2   & $2\times 10^{-21}$  & $2\times 10^{-21}$  \\
   $\rm V_c$     & $1{,}024^3$  & ABC       & 2    & 1      & 2   & $4\times 10^{-25}$  & $4\times 10^{-25}$   \\ 
   $\rm V_d$    & $2{,}048^3$  & ABC        & 2    & 1      & 2   & $1\times 10^{-27}$  & $1\times 10^{-27}$  \\
\hline
\end{tabular}
\label{Tbl:DetailedParameters}
\end{table*}

\section{Helical forcing results}
The section shows numerical results from simulations with  helical forcing. The $\alpha$-effect of dynamo theory is believed to be important in helical turbulence. We show here the helically forced results for completeness, although neglecting the $\alpha$ term in eq. \eqref{eq:pi_b_eta} may not be justified. Nevertheless, our results are remarkably similar to those in the main text.

\begin{figure*}[!htb]
\gridline{\fig{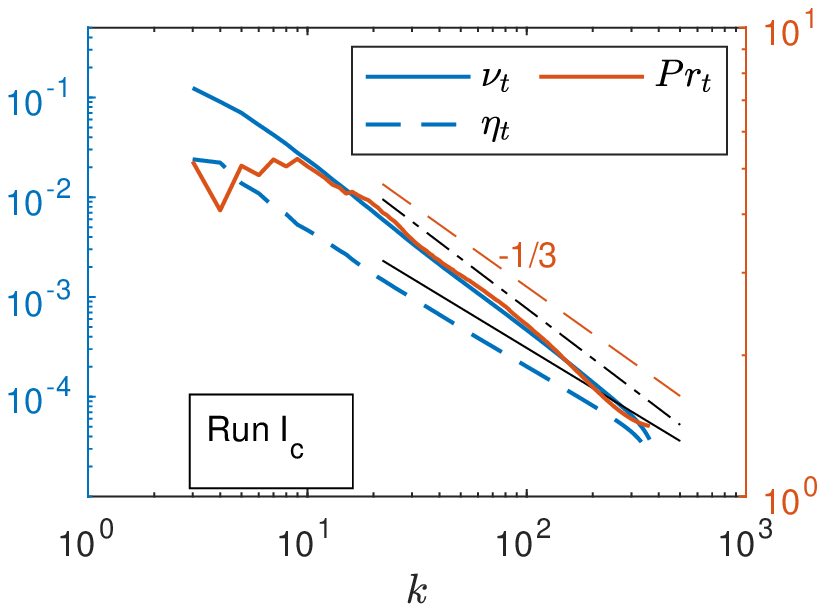}{0.3\textwidth}{}
          \fig{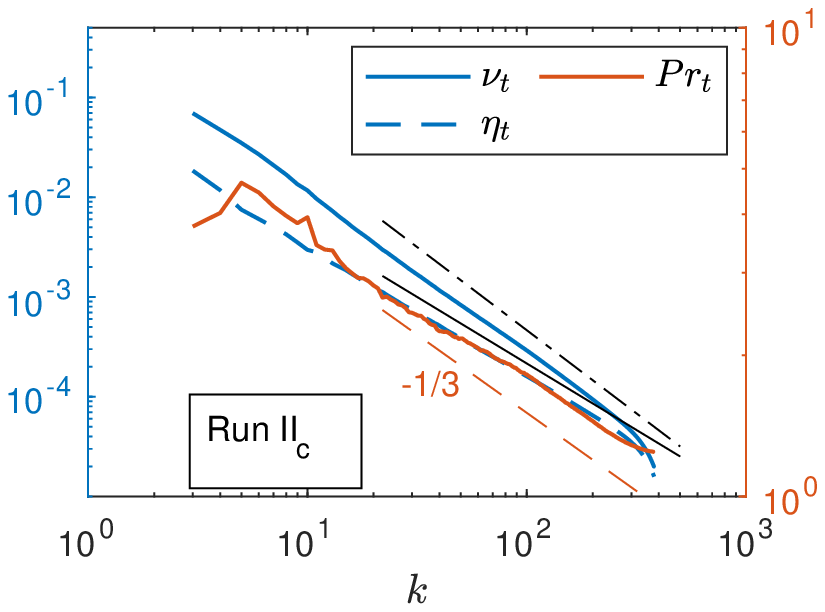}{0.3\textwidth}{}
          \fig{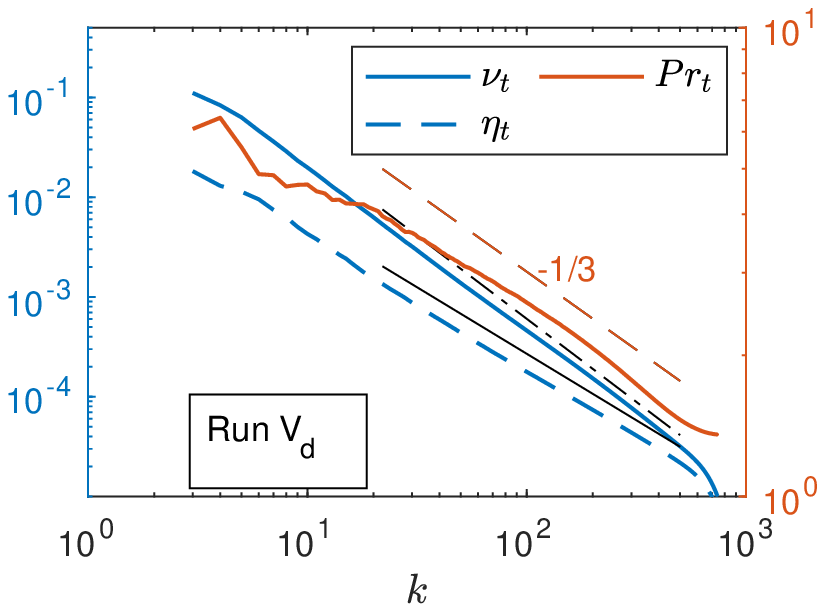}{0.3\textwidth}{}
          }
\caption{Plots showing turbulent viscosity $\nu_t$, turbulent resistivity $\eta_t$, and turbulent magnetic Prandtl number $Pr_t$ calculated using their respective definitions in eqs. \eqref{eq:pi_u_nu}-\eqref{eq:pr_def}, at different scales $k = L/\ell$. We use the highest resolution runs of Run I, II, and V (ABC forcing) in Table \ref{Tbl:Simulations}. Three reference lines with a slope of -1/3, -5/3 (black dash-dotted), and -4/3 (black solid) are added. Note the reference line of -1/3 and $Pr_t$ use the RIGHT $y$-axis, while others use the LEFT $y$-axis. Scales $<\ell_d$ are not shown.}
\label{fig:vis_res_pr-abc}
\end{figure*}

Figure \ref{fig:vis_res_pr-abc} shows $\nu_t$, $\eta_t$, and $Pr_t$ scaling at the highest resolution in helical forcing simulations, as a supplement to Fig. \ref{fig:vis_res_pr}(a)(b). The results are $\nu_t(k)\sim k^{-5/3}$ (or $\sim k^{-3/2}$) and $\eta_t(k)\sim k^{-4/3}$, and $Pr_t(k)\sim k^{-1/3}$ (or $\sim k^{-1/6}$), similar to the non-helical simulation results. As we mention in the main section, $\sigma_u$ is  $\approx1/4$ rather than 1/6  in the presence of a strong external B-field (Run $\rm II$), leading to the change in the scaling of $\nu_t$ and $Pr_t$.

\begin{figure*}[!ht]
\gridline{\fig{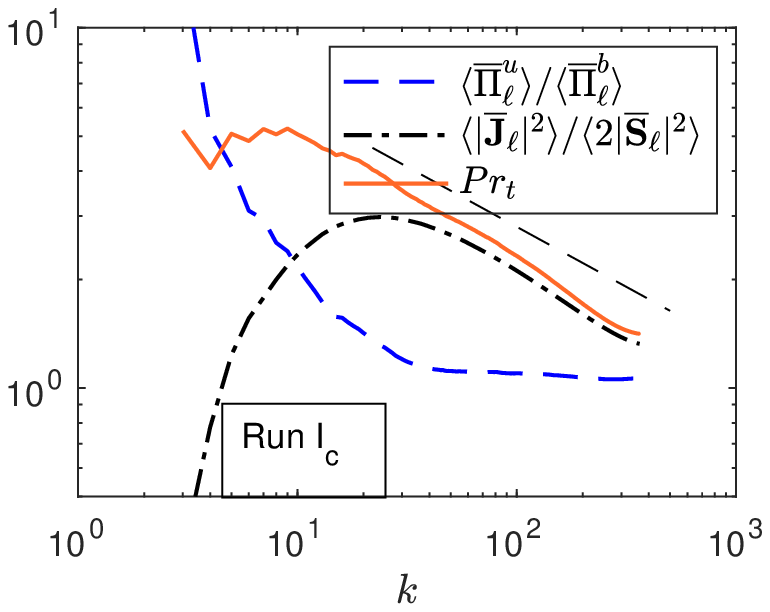}{0.3\textwidth}{}
          \fig{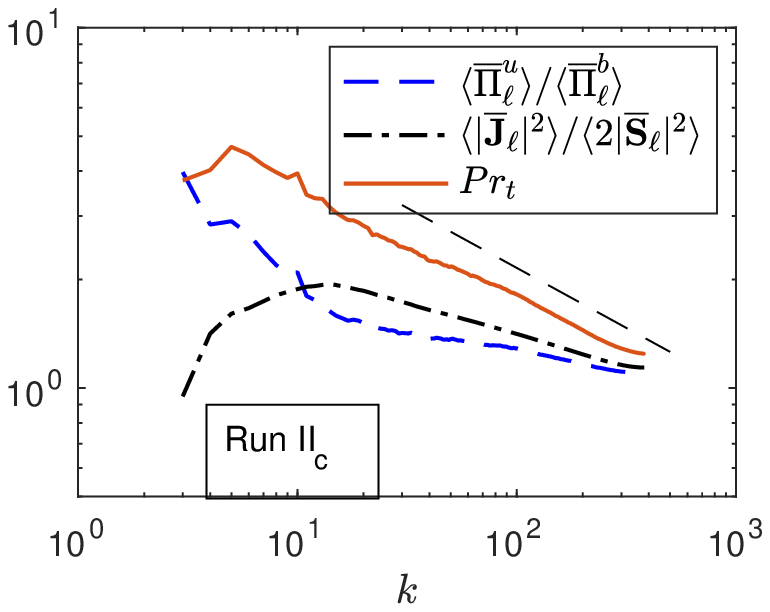}{0.3\textwidth}{} \fig{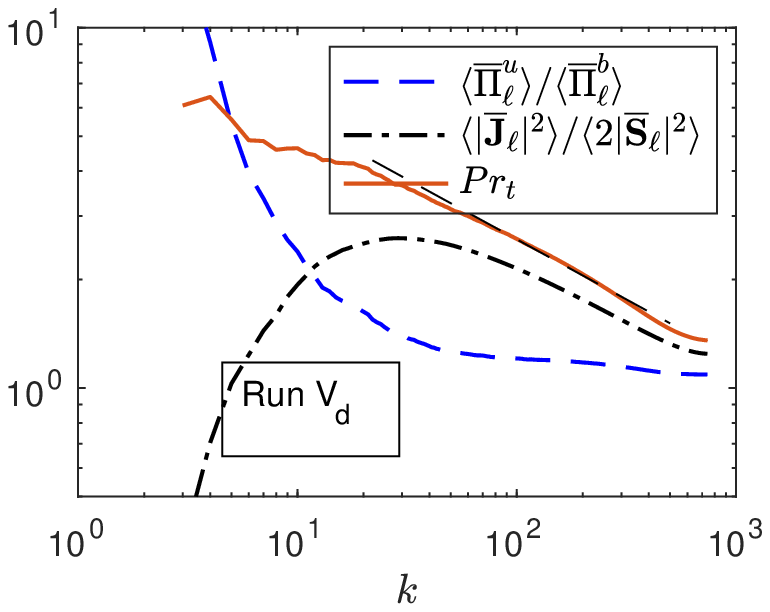}{0.3\textwidth}{}
          }
\caption{Plots showing the turbulent magnetic Prandtl number $Pr_t$ and its two components $\langle \OL \Pi^u_\ell \rangle / \langle \OL \Pi^b_\ell \rangle$ and $\langle |\OL \bJ_\ell|^2 \rangle/ \langle 2|\OL \bS_\ell|^2\rangle$ at the highest resolutions of Run I, II, and V (ABC forcing) in Table \ref{Tbl:Simulations}. A reference line with a slope of -1/3 (black dashed) is added. The plots show that $\langle \OL \Pi^u_\ell \rangle / \langle \OL \Pi^b_\ell \rangle$ approaches a constant in the decoupled range. Note that with a strong external B-field (Run II), we expect $\langle \OL \Pi^u_\ell \rangle / \langle \OL \Pi^b_\ell \rangle$ to plateau at sufficiently high Reynolds numbers \citep{bian2019decoupled}.}
\label{fig:pi_spec_pr-abc}
\end{figure*}

Figure \ref{fig:pi_spec_pr-abc} shows $\langle \OL \Pi^u_\ell \rangle / \langle \OL \Pi^b_\ell \rangle$ and $\langle |\OL \bJ_\ell|^2 \rangle/  \langle 2|\OL \bS_\ell|^2\rangle$ at the highest resolution in helical forcing simulations, as a supplement to Fig. \ref{fig:pi_spec_pr}. The results suggest constant  $\langle \OL \Pi^u_\ell \rangle / \langle \OL \Pi^b_\ell \rangle$ in the decoupled range and the 
same scaling of $\langle |\OL \bJ_\ell|^2 \rangle/  \langle 2|\OL \bS_\ell|^2\rangle$ and $Pr_t$ in the decoupled range, similar to the non-helical simulation results. The scaling of $\langle |\OL \bJ_\ell|^2 \rangle/  \langle 2|\OL \bS_\ell|^2\rangle$ is explained in the main section.

\begin{figure*}[!ht]
\gridline{\fig{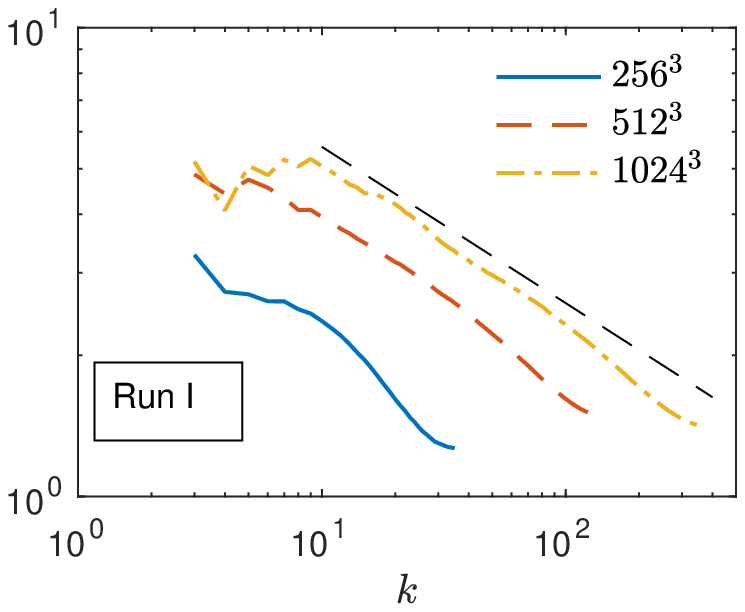}{0.3\textwidth}{}
          \fig{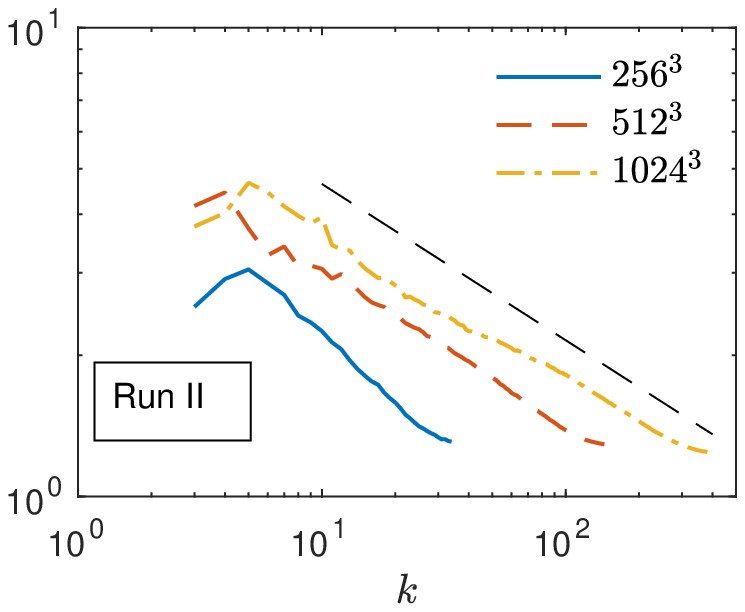}{0.3\textwidth}{}
          \fig{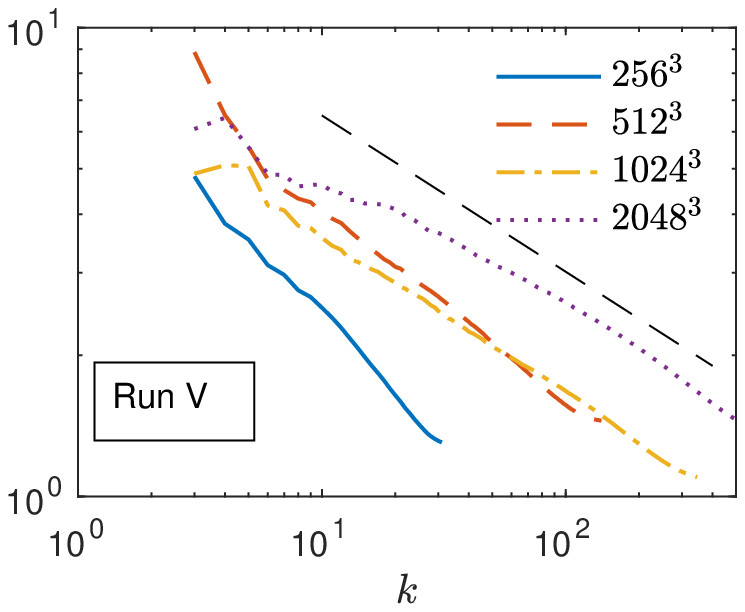}{0.3\textwidth}{}
          }
\caption{Plots showing the turbulent magnetic Prandtl number $Pr_t$ at different Reynolds numbers (grid resolution) of Run I, II, and V (ABC forcing) in Table \ref{Tbl:Simulations}. A reference line with a slope of -1/3 (black dashed) is added. }
\label{fig:Pr_diffRes_unnorm-abc}
\end{figure*}

Figure \ref{fig:Pr_diffRes_unnorm-abc} shows $Pr_t$ at different Reynolds number in helical forcing simulations, as a supplement to Fig. \ref{fig:Pr_diffRes_unnorm}. The results are similar to non-helical simulation results.

\section{Results at different Reynolds numbers}

\begin{figure*}
\gridline{\fig{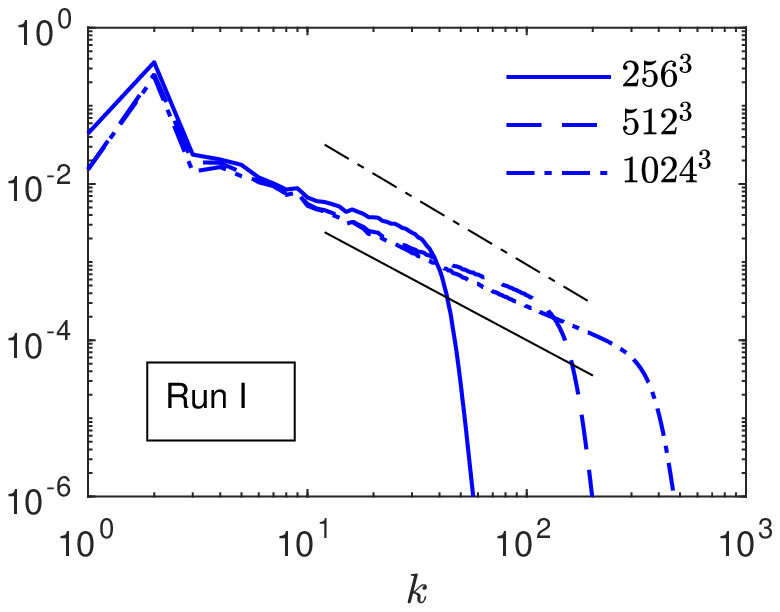}{0.3\textwidth}{}
          \fig{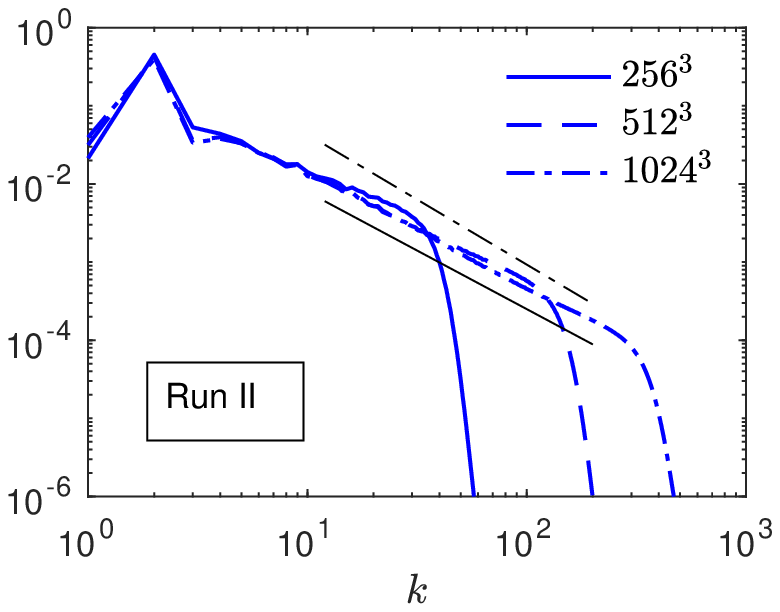}{0.3\textwidth}{}
          }
\gridline{ \fig{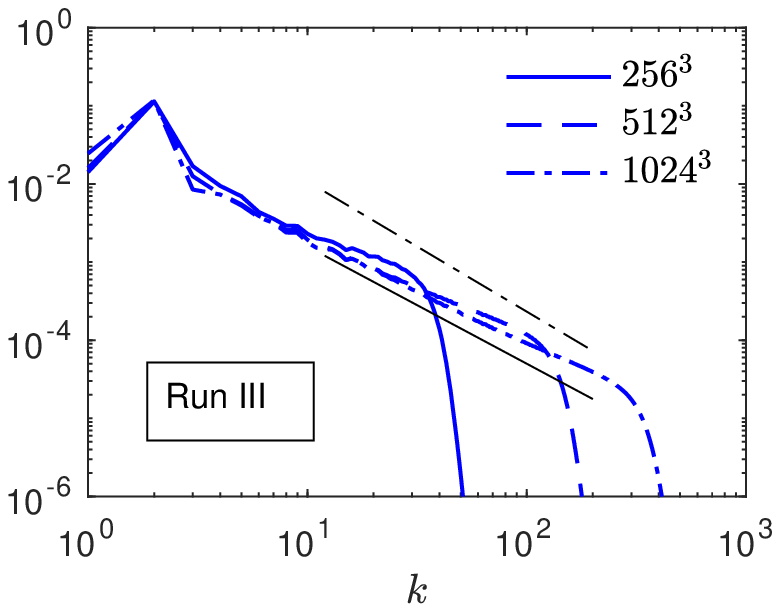}{0.3\textwidth}{}
           \fig{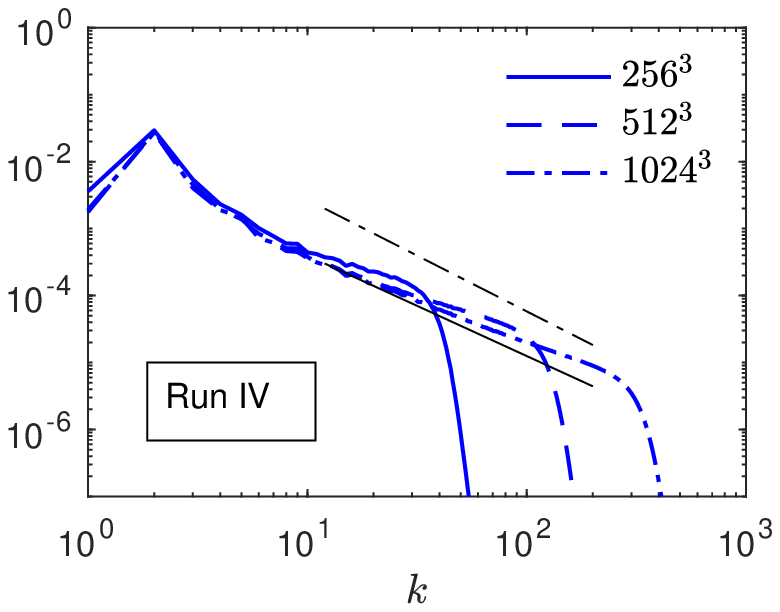}{0.3\textwidth}{}
          \fig{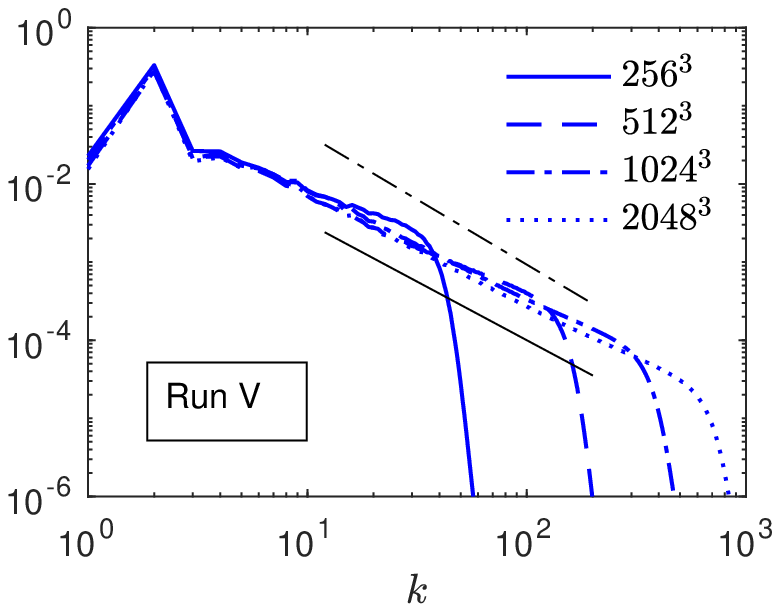}{0.3\textwidth}{}
          }
\caption{Plots showing kinetic energy spectrum $E^u(k)$ at different Reynolds numbers (grid resolution) of Run I-V. We show two reference lines with a slope of -3/2 (black solid), -5/3 (black dash-dotted). The slope becomes steeper as Reynolds number increases. The slop is near -3/2 at the highest resolution in Run II and Run V (with external B-field), while shallower than -3/2 in other simulations. Note \cite{grete2020matter} observed a kinetic energy spectrum of -4/3, which is also shallower than -3/2.}
\label{fig:spec_u}
\end{figure*}

Figure \ref{fig:spec_u} shows the kinetic energy spectrum at different Reynolds numbers (grid resolution). The slope becomes steeper as Reynolds number increases. The slop is near -3/2 at the highest resolution in Run II and Run V (with external B-field), while shallower than -3/2 in other simulations. \cite{grete2020matter} observed a kinetic energy spectrum of -4/3, which is also shallower than -3/2.

\begin{figure*}
\gridline{\fig{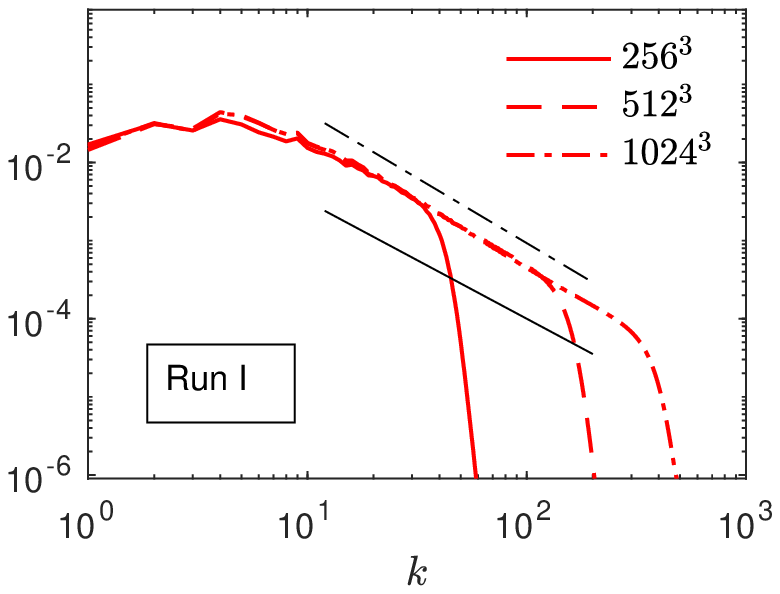}{0.3\textwidth}{}
          \fig{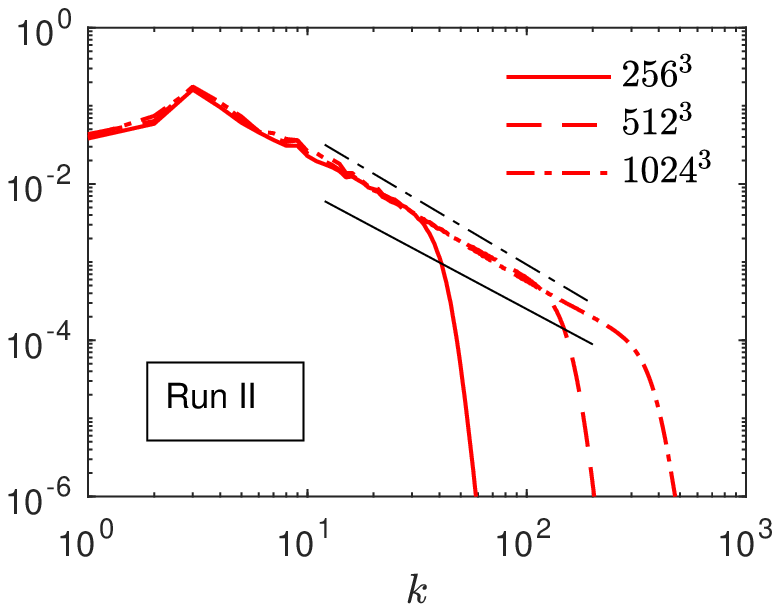}{0.3\textwidth}{}
          }

\gridline{ \fig{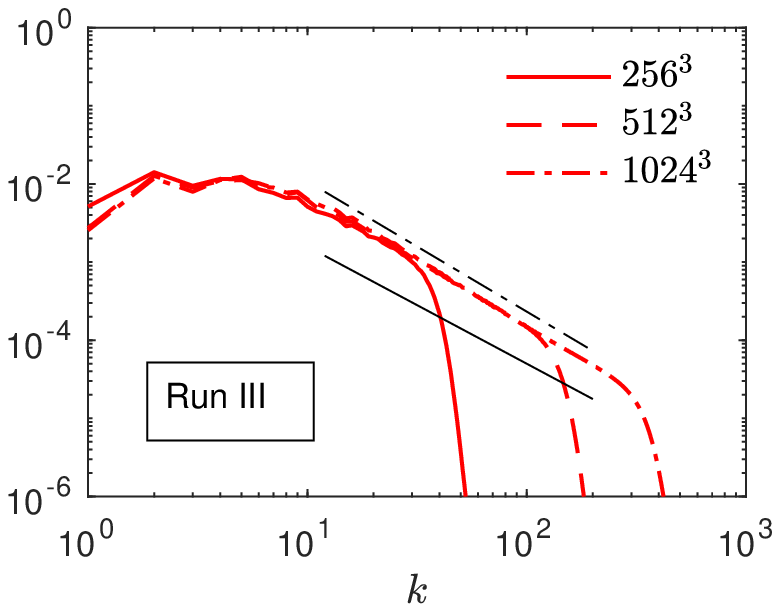}{0.3\textwidth}{}
           \fig{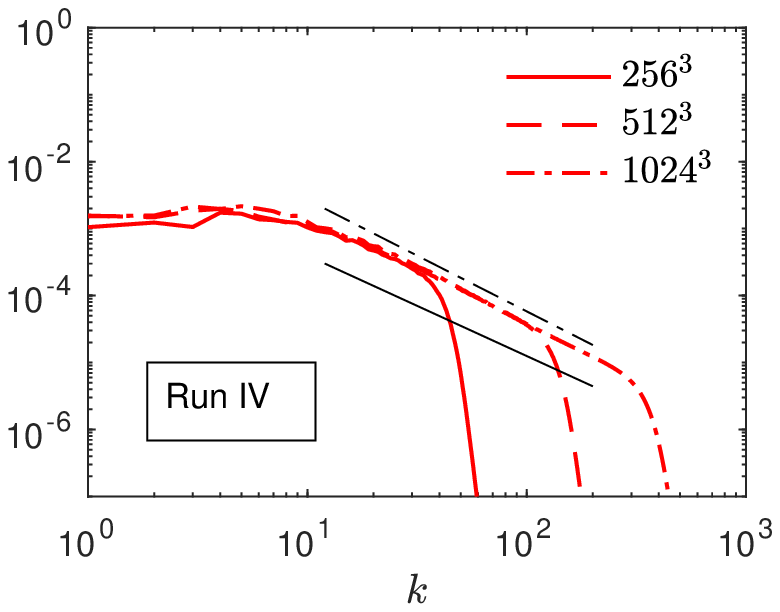}{0.3\textwidth}{}
          \fig{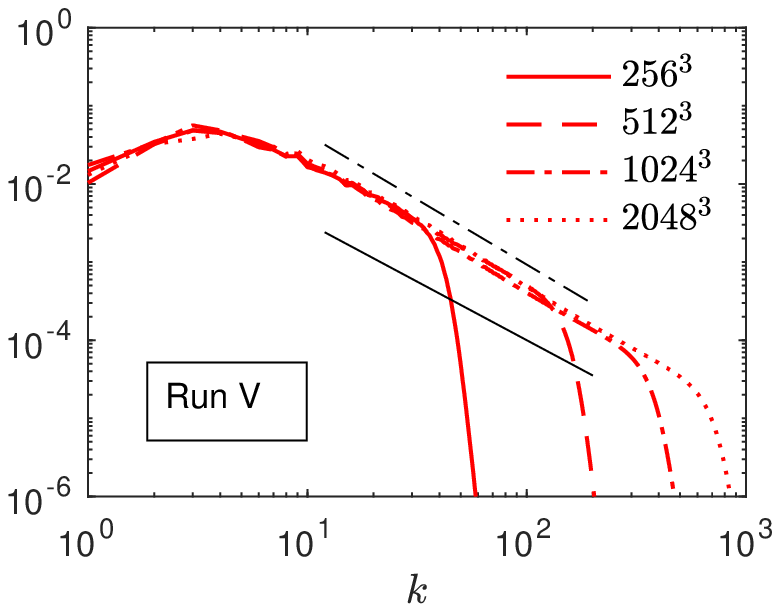}{0.3\textwidth}{}
          }
\caption{Plots showing magnetic energy spectrum $E^b(k)$ at different Reynolds numbers (grid resolution) of Run I-V. We show two reference lines with slope of -3/2 (black solid), -5/3 (black dash-dotted). The slope of magnetic spectrum agrees well with solar wind observations (-5/3).}
\label{fig:spec_b}
\end{figure*}

\begin{figure*}
\gridline{ \fig{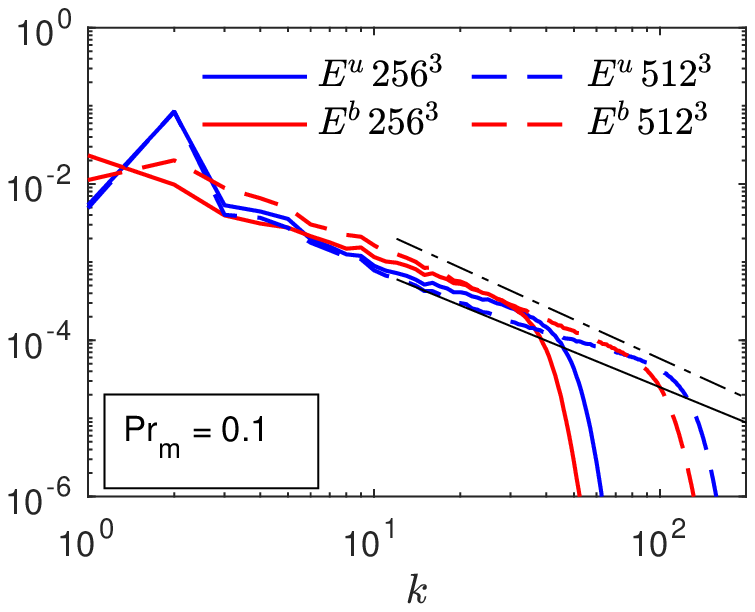}{0.3\textwidth}{}
           \fig{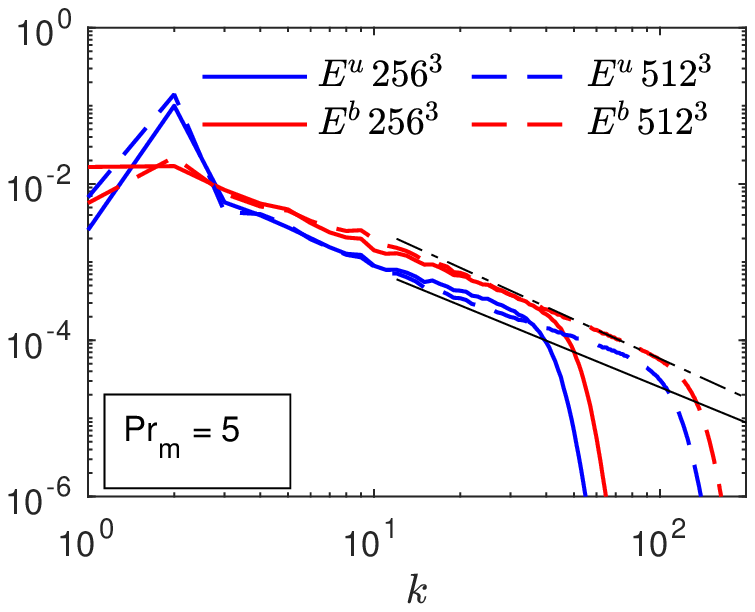}{0.3\textwidth}{}
          \fig{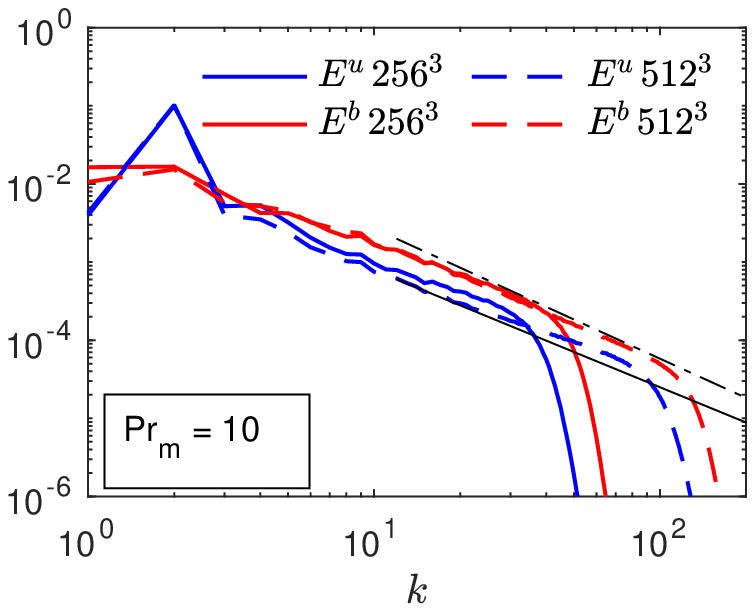}{0.3\textwidth}{}
          }
\caption{Plots showing kinetic and magnetic energy spectra, $E^u(k)$ and $E^b(k)$, of Run IV with $Pr_m$ = 0.1, 5, and 10. We show two reference lines with slope of -3/2 (black solid), -5/3 (black dash-dotted).}
\label{fig:spec_nonunity}
\end{figure*}

Figure \ref{fig:spec_b} shows the magnetic energy spectrum at different Reynolds numbers. The slope agrees well with $-5/3$ for all Reynolds numbers. Figure \ref{fig:spec_nonunity} shows the kinetic and magnetic energy spectra of Run IV with $Pr_m$ = 0.1, 5, and 10. 

\begin{figure*}
\gridline{\fig{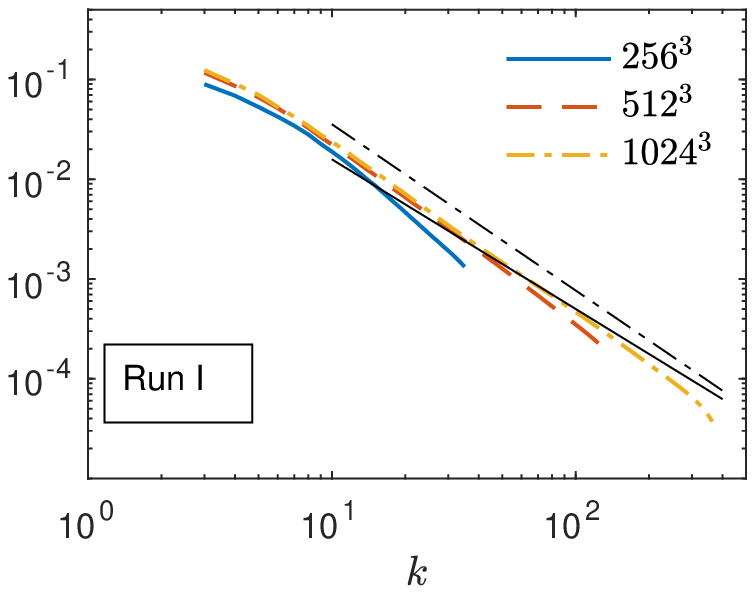}{0.3\textwidth}{}
          \fig{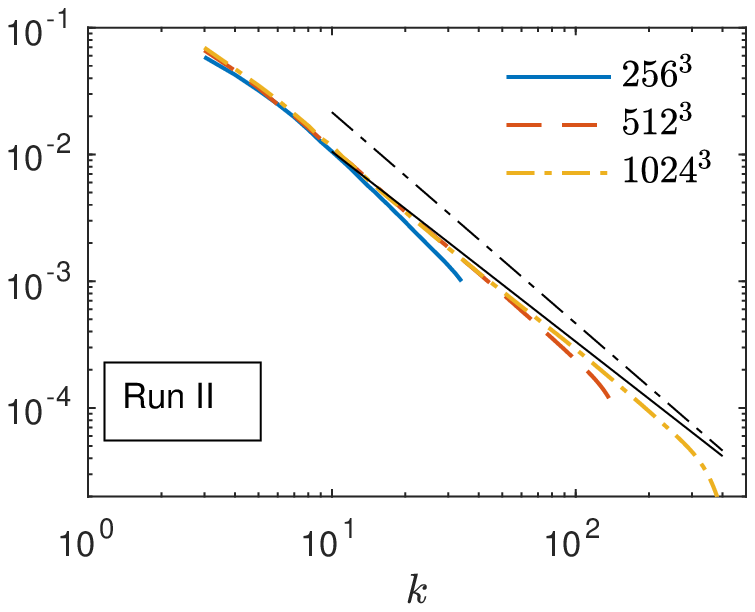}{0.3\textwidth}{}
          }
\gridline{\fig{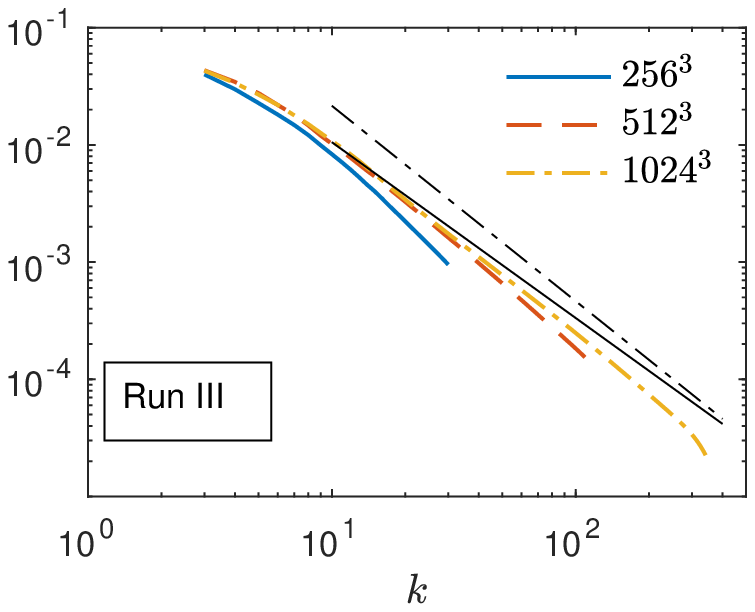}{0.3\textwidth}{}
          \fig{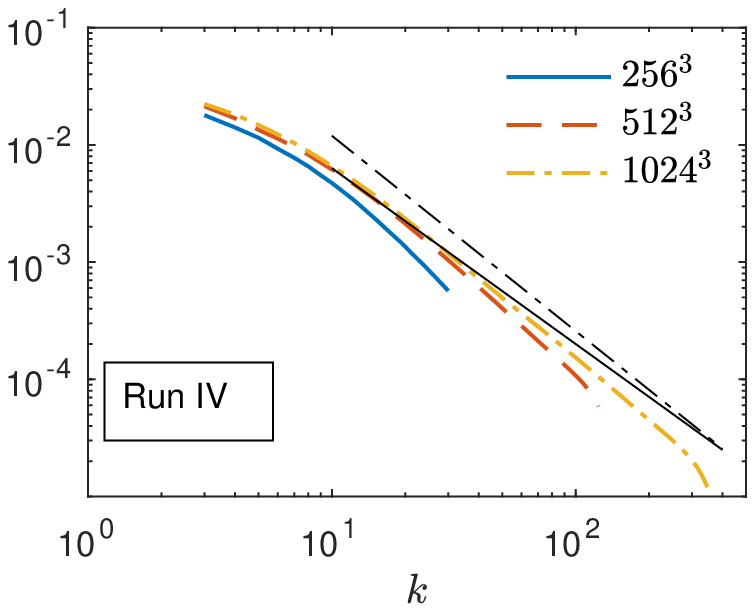}{0.3\textwidth}{}
          \fig{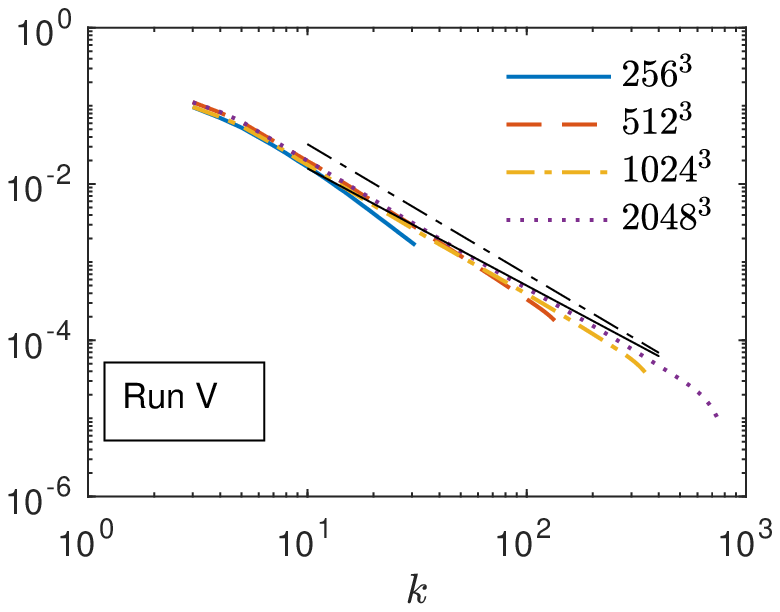}{0.3\textwidth}{}
          }
\caption{Plots showing turbulent viscosity $\nu_t$ at different Reynolds numbers (grid resolution) of Run I-V. Reference lines with slope of -5/3 (black dash-doted) and -3/2 (black solid) are added.}
\label{fig:nu_diffRes}
\end{figure*}

\begin{figure*}
\gridline{\fig{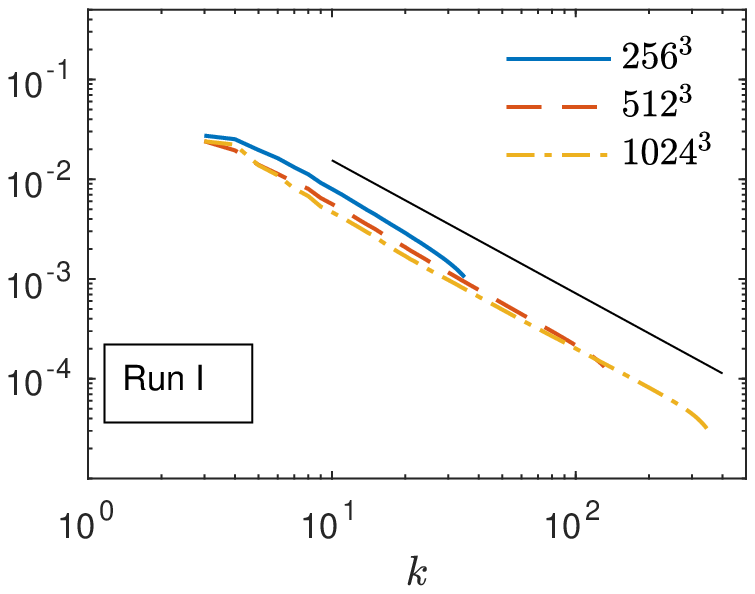}{0.3\textwidth}{}
          \fig{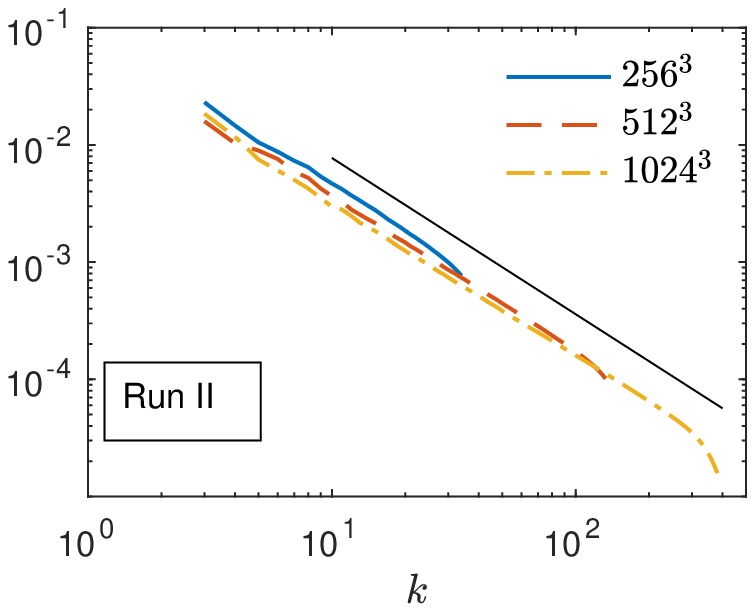}{0.3\textwidth}{}
          }
\gridline{\fig{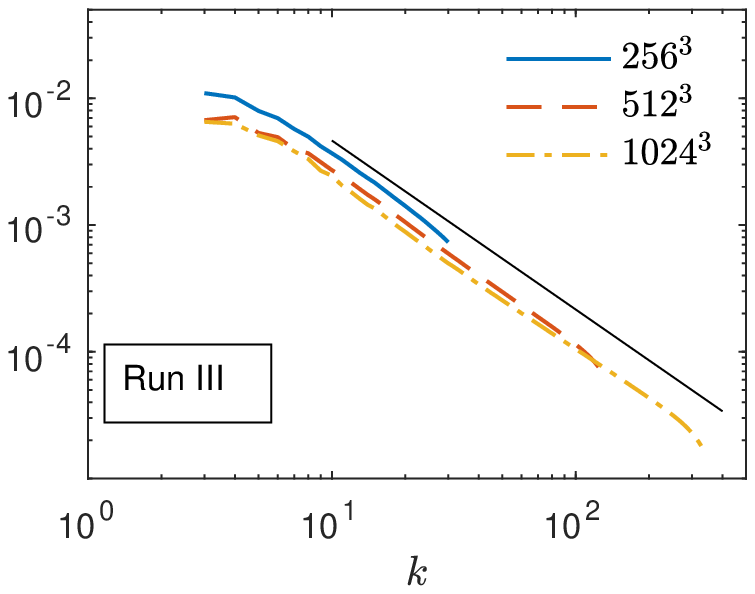}{0.3\textwidth}{}
          \fig{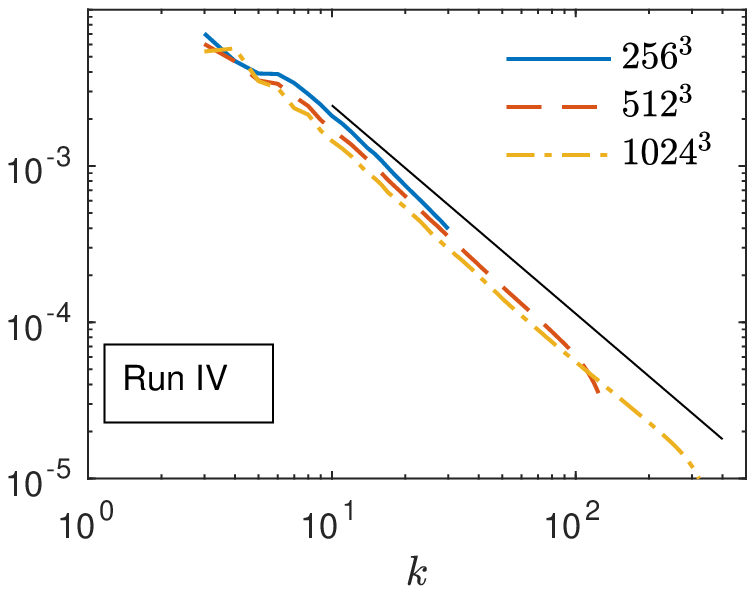}{0.3\textwidth}{}
          \fig{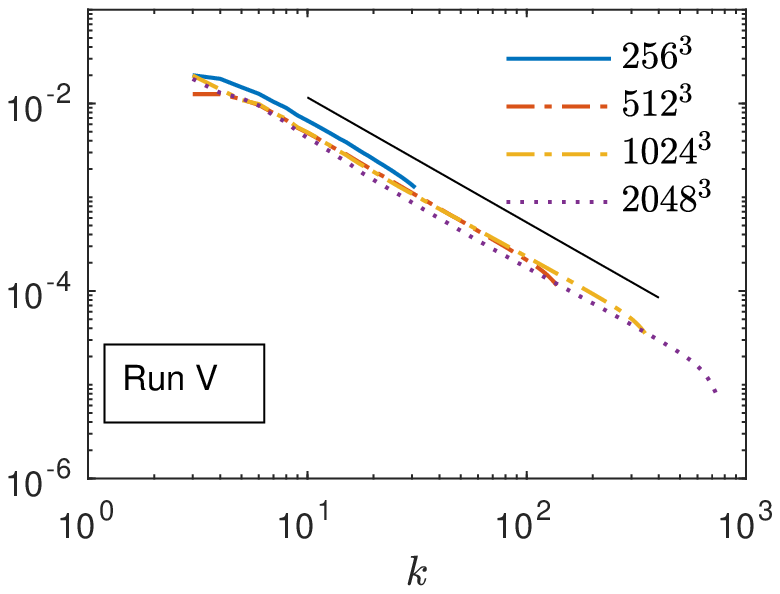}{0.3\textwidth}{}
          }
\caption{Plots showing turbulent resistivity of $\eta_t$ at different Reynolds numbers (grid resolution) of Run I-V. A reference line with slope of -4/3 (black solid) is added. The scaling exponent agrees well with the expected value -4/3.}
\label{fig:eta_diffRes}
\end{figure*}

Figure \ref{fig:nu_diffRes} shows that the scaling exponent of $\nu_t$ is near $-5/3$ ($-3/2$ in Run $\rm II_c$ and $\rm V_d$) at the highest resolution. As Reynolds number increases, it becomes steeper and approaches -3/2. Figure \ref{fig:eta_diffRes} shows the scaling exponent of $\eta_t$ at all Reynolds numbers is near $-4/3$, consistent with eq. \eqref{eq:scaling_solar}. 

\begin{figure*}
\gridline{\fig{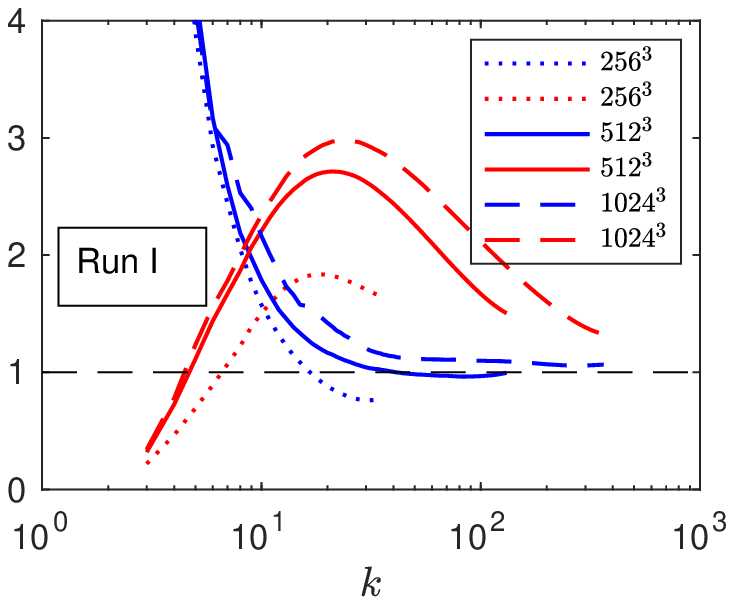}{0.3\textwidth}{}
          \fig{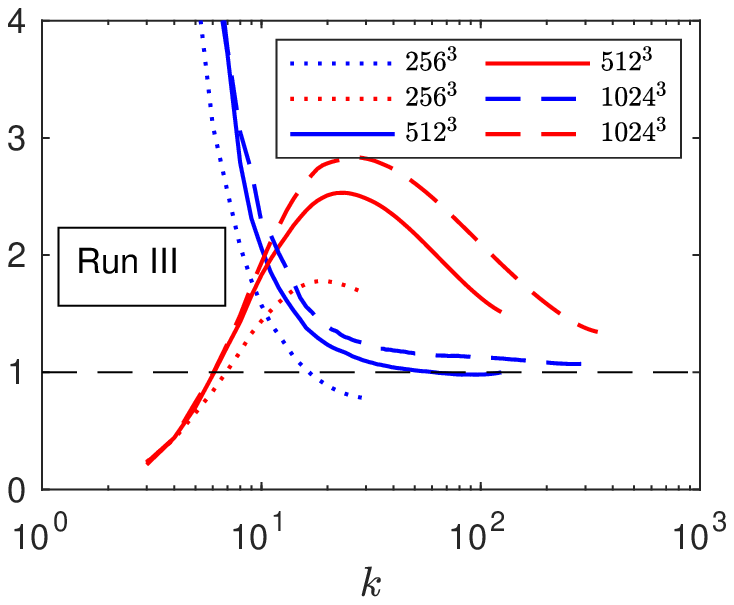}{0.3\textwidth}{}
          \fig{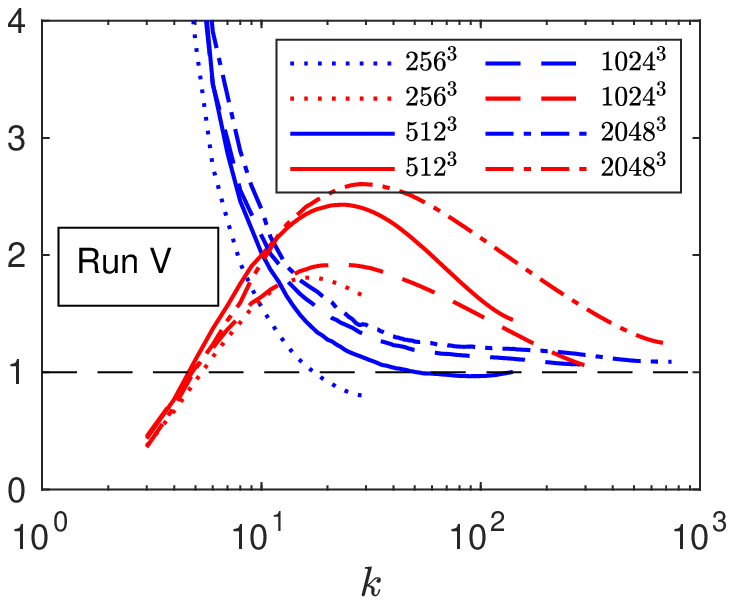}{0.3\textwidth}{}
          }
\gridline{\fig{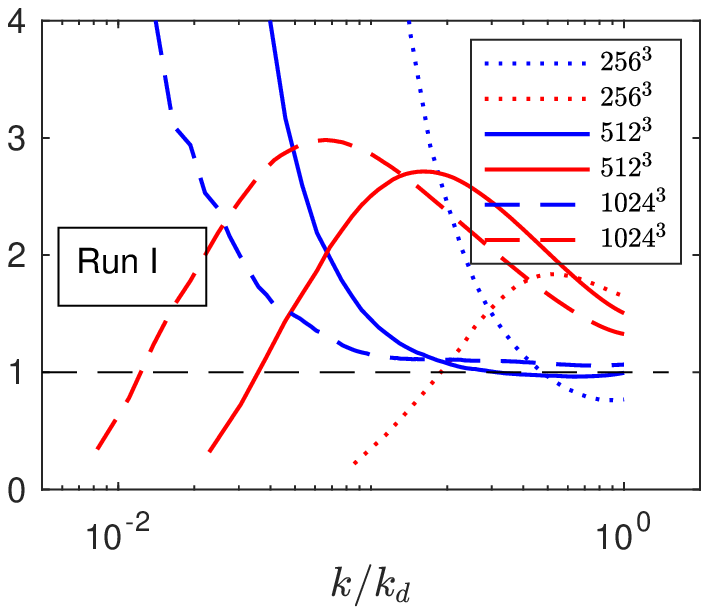}{0.3\textwidth}{}
        \fig{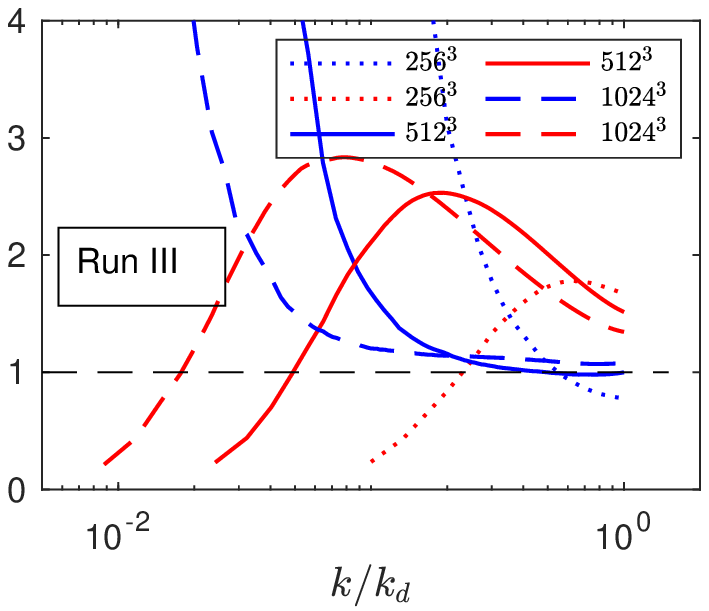}{0.3\textwidth}{}
          \fig{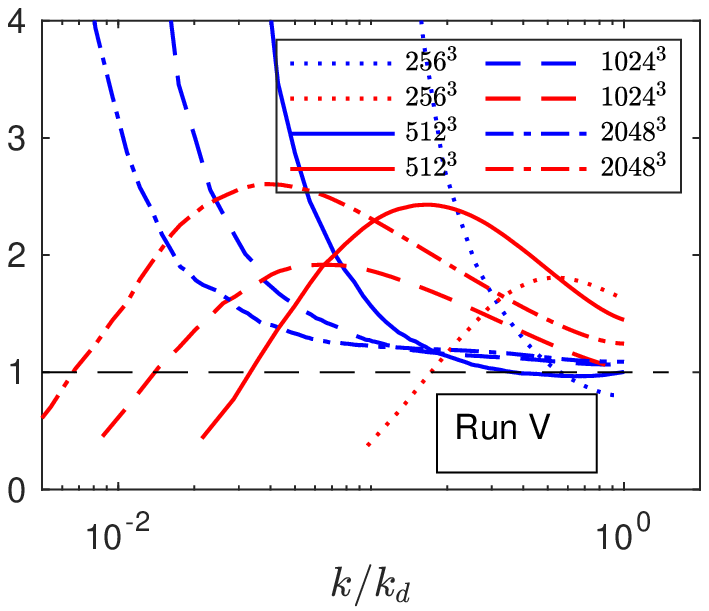}{0.3\textwidth}{}
          }
\caption{Plots showing $\langle \OL \Pi^u_\ell \rangle / \langle \OL \Pi^b_\ell \rangle$ (blue lines) and $\langle |\OL \bJ_\ell|^2 \rangle/ \langle 2|\OL \bS_\ell|^2\rangle$ (red lines) at different Reynolds numbers (grid resolution) of Run I, III and V. The $x$-axis in bottom panels is normalized by $k_d=L/\ell_d$. A reference line (black dashed) of 1 is added.}
\label{fig:pi_js_ratio}
\end{figure*}

Figure \ref{fig:pi_js_ratio} shows that $\langle \OL \Pi^u_\ell \rangle / \langle \OL \Pi^b_\ell \rangle$ and $\langle |\OL \bJ_\ell|^2 \rangle/ \langle 2|\OL \bS_\ell|^2\rangle$ at different Reynolds numbers (grid resolution). 

\begin{figure*}
\gridline{\fig{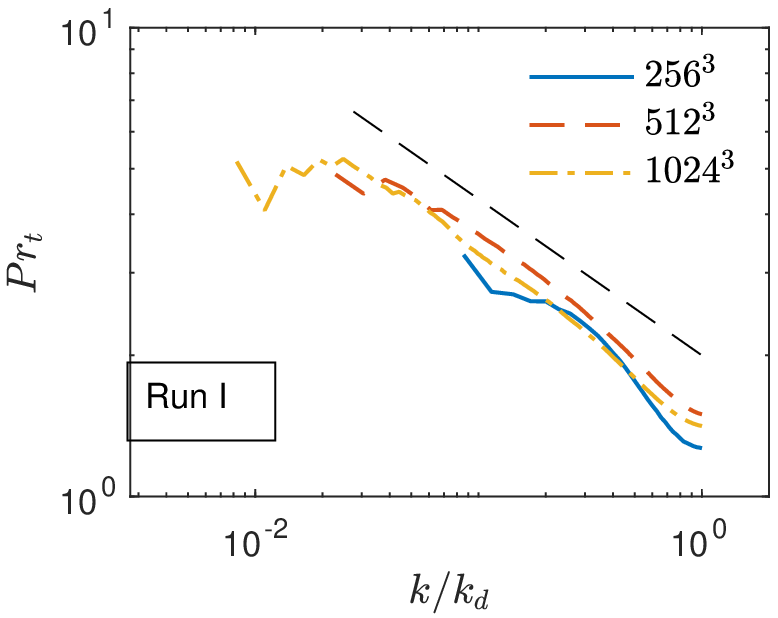}{0.3\textwidth}{}
          \fig{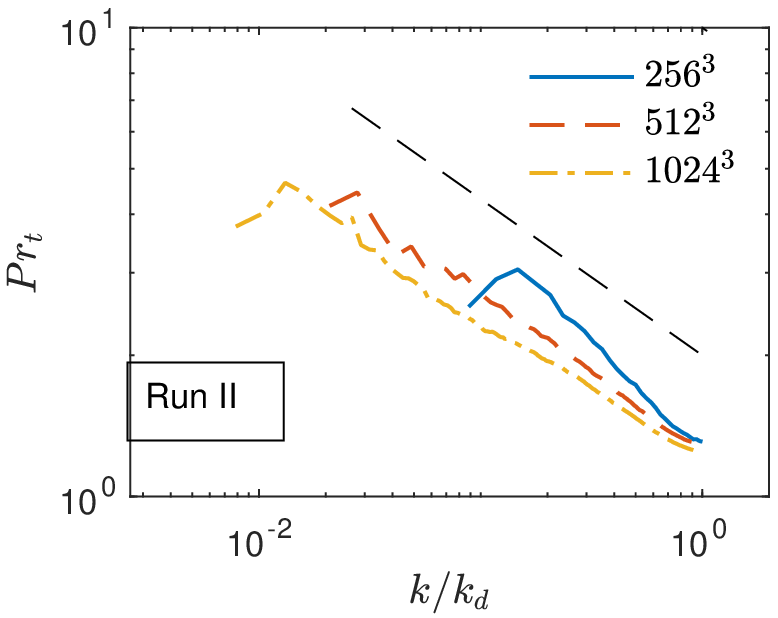}{0.3\textwidth}{}
          }
\gridline{ \fig{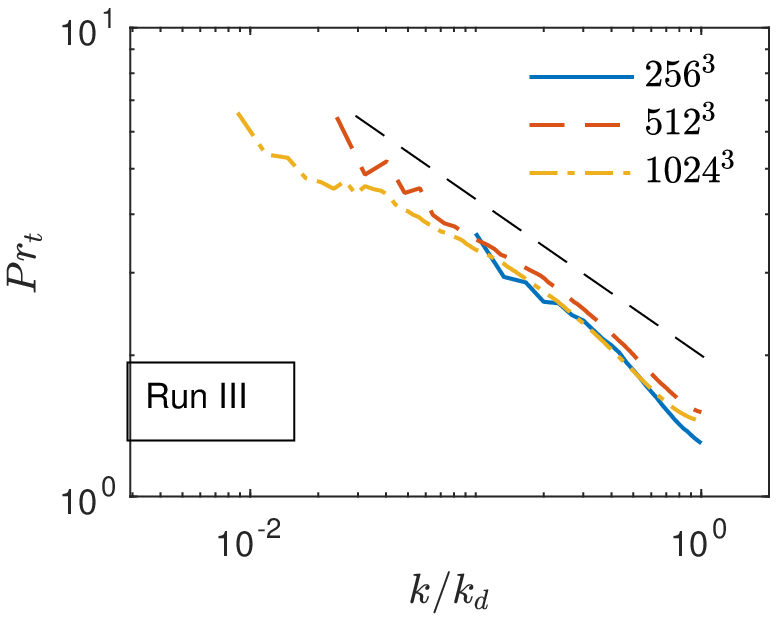}{0.3\textwidth}{}
 \fig{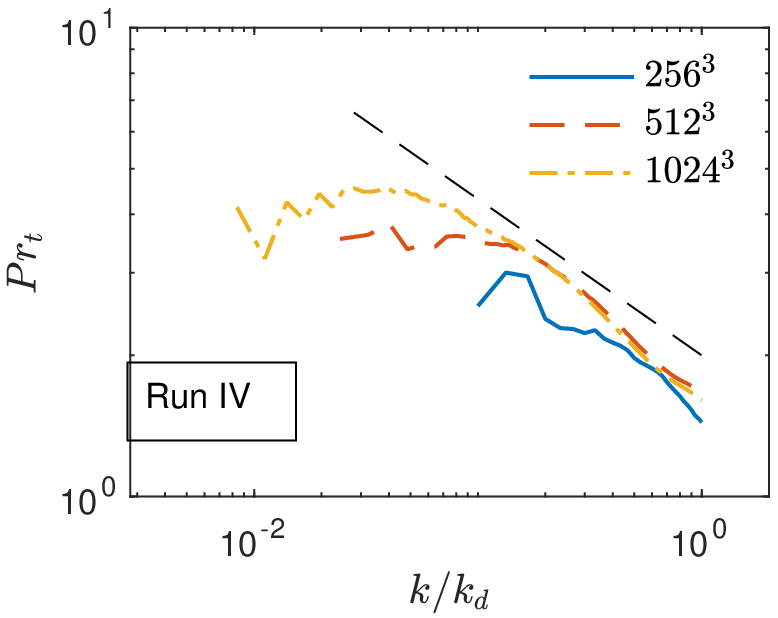}{0.3\textwidth}{}
          \fig{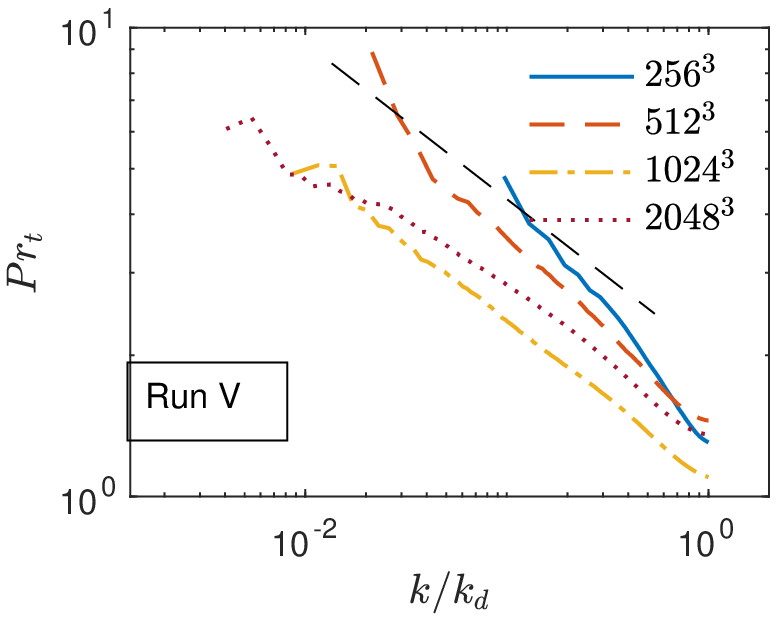}{0.3\textwidth}{}
          }
\caption{Plots showing $Pr_t$ at different Reynolds numbers (grid resolution) of Run I-V with $x$-axis normalized by $k_d=2\pi/\ell_d$. A reference line with a slope of -1/3 (black dashed) is added. $Pr_t$ at different Reynolds numbers collapse at $k=k_d$, as expected (see also Table \ref{Tbl:Pr_t_kd}).}
\label{fig:Pr_diffRes_norm}
\end{figure*}

Figure \ref{fig:Pr_diffRes_norm} shows $Pr_t$ at different Reynolds numbers with $x$-axis normalized by $k_d=L/\ell_d$, where $\ell_d$ is defined as the scale at which $ \langle\OL \Pi^u_{\ell} + \OL\Pi^b_{\ell}\rangle = \nu \langle|\nabla\OL{\textbf u}|^2\rangle + \eta \langle|\nabla\OL{\textbf B}|^2\rangle$. For non-unity $Pr_m$, $\ell_d\equiv \text{max} (\ell_{\nu}, \ell_{\eta})$. $\ell_{\nu}$ and $\ell_\eta$ are defined as scales where $ \langle\OL \Pi^u_\ell\rangle = \nu \langle|\nabla\OL{\textbf u}|^2\rangle$ and $\langle \OL\Pi^b_\ell\rangle =  \eta \langle|\nabla\OL{\textbf B}|^2\rangle$. $Pr_t$ at different Reynolds numbers collapse at $k=k_d$, as expected (see also Table \ref{Tbl:Pr_t_kd}).

\begin{figure}[!ht]
\gridline{\fig{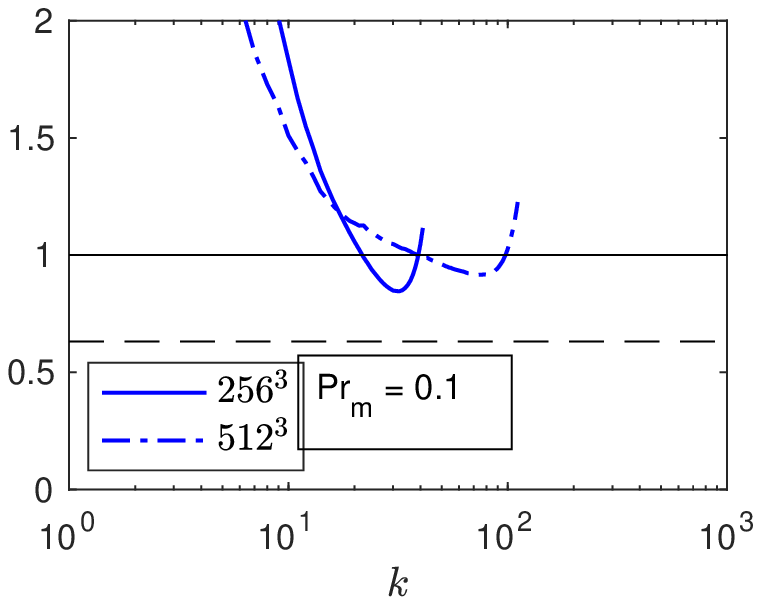}{0.3\textwidth}{}
          \fig{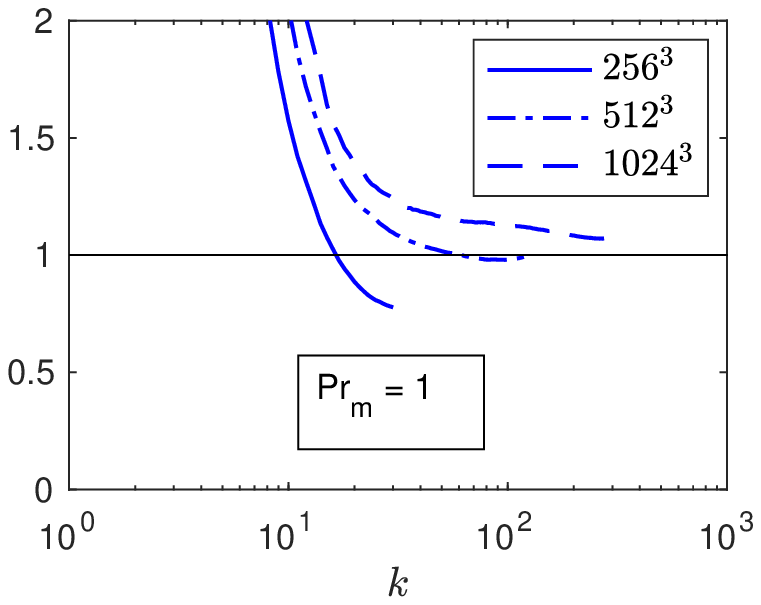}{0.3\textwidth}{}
          }
\gridline{\fig{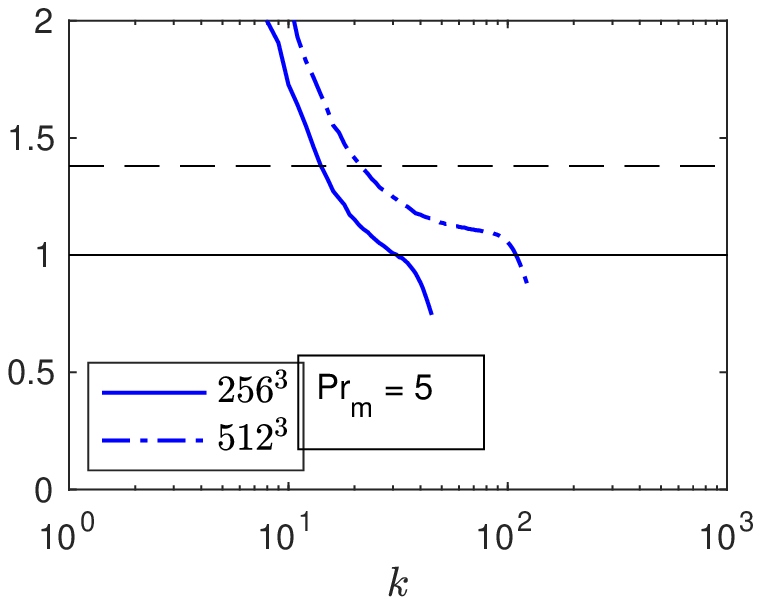}{0.3\textwidth}{}
          \fig{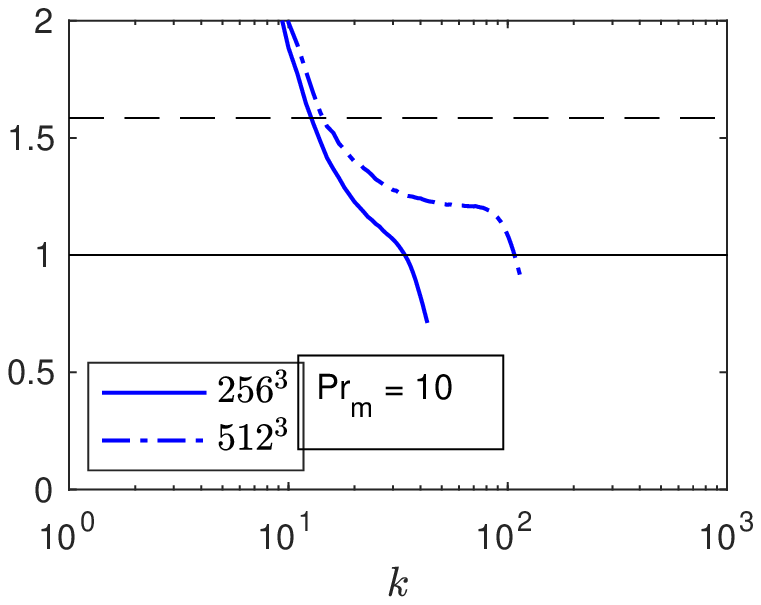}{0.3\textwidth}{}
          }
\caption{Plots showing $\langle \OL \Pi^u_\ell \rangle / \langle \OL \Pi^b_\ell \rangle$ at different microscopic Prandtl numbers ($Pr_m$ = 0.1, 1, 5, 10). The parameters are detailed in Table \ref{Tbl:DetailedParameters}. These simulations are conducted on $512^3$ grid. A reference line (black solid) of 1 is added in all panels. Note our usage of fifth-order hyperdiffusion in the simulations. Another reference line (black dashed) of  $\nu_h^{1/5}/\eta_h^{1/5}$ is added as an estimate for the microscopic magnetic Prandtl number corresponding to normal (Laplacian) diffusion. The estimate is 0.63, 1, 1.38, and 1.58 for $Pr_m=0.1$, $Pr_m=1$, $Pr_m=5$, and $Pr_m=10$, respectively. Since the decoupled range, over which $\langle \OL \Pi^u_\ell \rangle$ and $\langle \OL \Pi^b_\ell \rangle$ become scale-independent, is barely resolved, these plots neither reinforce nor conflict with the expectation of asymptotic equipartition of the kinetic and magnetic cascades predicted in \cite{bian2019decoupled}, irrespective of microscopic $Pr_m$.}
\label{fig:cascade_ratio_different_pr}
\end{figure}

Figure \ref{fig:cascade_ratio_different_pr} shows $\langle \OL \Pi^u_\ell \rangle / \langle \OL \Pi^b_\ell \rangle$ at different microscopic Prandtl numbers ($Pr_m$ = 0.1, 1, 5, 10). Since the decoupled range, over which each of $\langle \OL \Pi^u_\ell \rangle$ and $\langle \OL \Pi^b_\ell \rangle$ becomes scale-independent, is barely resolved, these plots neither reinforce nor conflict with the expectation of asymptotic equipartition of the kinetic and magnetic cascades predicted in \cite{bian2019decoupled}, irrespective of microscopic $Pr_m$. It is worth emphasizing that the observation of \cite{brandenburg2014magnetic} of a positive correlation between $Pr_m$ and the ratio of kinetic dissipation to magnetic dissipation, does not have a direct bearing on the ratio of the cascades. This is because the cascades $\langle \OL \Pi^u_\ell \rangle$ and $\langle \OL \Pi^b_\ell \rangle$ in the decoupled range are not necessarily equal to the kinetic and magnetic energy dissipation, respectively. This is especially true at non-unity $Pr_m$ at scales smaller than $\ell_d$ beyond the decoupled range, where kinetic-magnetic conversion is expected to occur (e.g., in the viscous-inductive range at high $Pr_m$) before all energy is dissipated microscopically.

\begin{table*}[!ht]
\centering
\caption{$Pr_t$ at $k/k_d=1$, where $k_d = L/\ell_d$.}
\begin{tabular}{llllll}
\hline
\hline
         & Run I                & Run II & Run III & Run IV & Run V \\ \hline
$256^3$  & 1.23                 & 1.31   & 1.40    & 1.44   & 1.30  \\
$512^3$  & 1.50                 & 1.29   & 1.45    & 1.68   & 1.45  \\
$1024^3$ & 1.41                 & 1.25   & 1.51    & 1.67   & 1.10  \\
$2048^3$ & \multicolumn{1}{c}{} &        &         &        & 1.36  \\ \hline
\end{tabular}
\label{Tbl:Pr_t_kd}
\end{table*}

\end{document}